\documentclass[lettersize,journal]{IEEEtran}
\usepackage{amsmath,amsfonts, amssymb, bm}
\usepackage{algorithmic}
\usepackage{algorithm}
\usepackage{array}
\usepackage[caption=false,font=normalsize,labelfont=sf,textfont=sf]{subfig}
\usepackage{booktabs}
\usepackage{textcomp}
\usepackage{stfloats}
\usepackage{url}
\usepackage{verbatim}
\usepackage{graphicx}
\usepackage{cite}
\newcommand{\Statex}{\item[]}
\usepackage{xcolor}

\begin{document}

\title{DBPnet: Damper Characteristics-Based Bayesian Physics-Informed Neural Network for Wheel Load Estimation}

\author{Tianyi Wang$^{1*}$, Tianyi Zeng$^{2*}$, Zimo Zeng$^{3}$, Feiyang Zhang$^{4}$, Yujin Wang$^{4}$, Xiangyu Li$^{1}$, \\ Yiming Xu$^{5}$, Sikai Chen$^{6}$, Junfeng Jiao$^{5}$, Christian Claudel$^{1}$, Xinbo Chen$^{4\dag}$%
\thanks{$^\dag$Corresponding author: Xinbo Chen.}%
\thanks{$^{*}$These authors contributed equally to this work.}%
\thanks{$^{1}$Department of Civil, Architectural, and Environmental Engineering, The University of Texas at Austin, Austin, TX 78712, USA.
 	{\tt\small bonny.wang@utexas.edu, xiangyu\_li@utexas.edu, christian.claudel@utexas.edu}}%
\thanks{$^{2}$School of Automation and Intelligent Sensing, Shanghai Jiao Tong University, Shanghai 200240, China.
 	{\tt\small zengtianyi@sjtu.edu.cn}}%
\thanks{$^{3}$College of Electrical Engineering, Zhejiang University, Hangzhou 310027, China.
 	{\tt\small zimozeng904@gmail.com}}%
\thanks{$^{4}$School of Automotive Studies, Tongji University, Shanghai 201804, China.
 	{\tt\small 2151166@tongji.edu.cn, 2510180@tongji.edu.cn, chenxinbo@tongji.edu.cn}}%
\thanks{$^{5}$School of Architecture, The University of Texas at Austin, Austin, TX 78712, USA.
 	{\tt\small yiming.xu@utexas.edu, jjiao@austin.utexas.edu}}%
\thanks{$^{6}$Department of Civil and Environmental Engineering, University of Wisconsin-Madison, Madison, WI 53706, USA.
 	{\tt\small sikai.chen@wisc.edu}}%
}

\markboth{Journal of \LaTeX\ Class Files,~Vol.~14, No.~8, August~2021}%
{Shell \MakeLowercase{\textit{et al.}}: A Sample Article Using IEEEtran.cls for IEEE Journals}


\maketitle

\begin{abstract}
Advanced driver assistance systems (ADAS) play an important role in modern automotive intelligence, significantly enhancing vehicle safety and stability. 
The performance of ADAS critically relies on accurate and reliable vehicle state estimation, particularly from vehicle dynamic sensors. 
Among these signals, wheel load is a key variable for chassis control and safety-critical functions, yet it remains difficult to estimate robustly due to complex suspension geometry, nonlinear dynamics, and measurement noise. 
To address this issue, we propose DBPnet, a Bayesian physics-informed neural network (PINN) with a physics-aware embedding module inspired by damper characteristics.
First, this paper presents a suspension linkage-level modeling (SLLM) approach that constructs a nonlinear instantaneous dynamic model by explicitly considering the complex geometric structure of the suspension. 
Building upon SLLM, Bayesian inference is integrated into the PINN to effectively cope with noise and uncertainty in the vehicle chassis system, thereby improving the model's robustness. 
Then, a physics-informed loss function is employed to ensure consistency with fundamental physical principles, while the damper characteristics-inspired embedding module extracts temporal variation features of input signals and incorporates them into each layer of the PINN, ensuring that physical observations guide the neural network without being constrained by fixed physical models.
Extensive evaluations on high-fidelity simulations and real-world experiments demonstrate that our DBPnet consistently achieves lower RMSE and MaxError than baseline methods. 
These results highlight the potential of our DBPnet to advance wheel load estimation and contribute to the development of more reliable ADAS actuator functions.
\end{abstract}


\section{Introduction}
\label{sec:introduction}

Accurate state estimation in noisy multi-input multi-output (MIMO) systems, particularly for vehicle chassis dynamics, is fundamental to the development of robust advanced driver assistance systems (ADAS) \cite{zeng2023wheel}. 
However, this task remains highly challenging due to imperfect physical models and highly complex, nonlinear relationships between system inputs and outputs \cite{zeng2024analysis}. 
Traditional Kalman filter (KF)-like methods, such as the extended Kalman filter (EKF) \cite{ribeiro2004kalman} and ensemble Kalman filter (EnKF) \cite{roth2017ensemble}, provide closed-form Bayesian updates to Gaussian posterior approximations and scale to high dimensions, while they sometimes fail to achieve accurate estimation in such real-world scenarios because extreme observations and sensor measurement errors may corrupt Gaussian measurement models.

To overcome these limitations, deep learning approaches that integrate domain knowledge have recently shown significant promise in improving estimation accuracy and robustness \cite{wang2024ftekfnet}.
Many MIMO systems, such as robotics \cite{Yang2023}, vehicles \cite{tan2023vehicle}, and transportation systems \cite{wang2025hlcg}, are typically governed by partial differential equations (PDEs), which yield high-dimensional, nonlinear dynamical models with hundreds of state variables. 
Physics-informed neural networks (PINNs) have emerged as a powerful paradigm for both forward (inference) and inverse (identification) problems by embedding governing physical laws into the network's loss functions \cite{Sun2024}.  
By leveraging partial observations and physical constraints, PINNs can infer unknown parameters and reconstruct system states, representing a significant advance in solving complex state estimation tasks in nonlinear dynamical systems \cite{raj2025deep}. 

Beyond encoding physics laws into loss functions, recent work \cite{wang2025physics} has explored physics-guided architecture designs that embed external physical information directly into neural networks. 
A notable example is feature-wise linear modulation (FiLM) \cite{perez2018film}, which conditions intermediate network activations through feature-wise affine transformations based on external data. 
Although originally proposed for visual reasoning, FiLM-inspired architectures provide an effective mechanism for injecting physical context from auxiliary sensors, thereby dynamically influencing the network's computational flow.
Despite the fact that PINNs have demonstrated success in integrating physics into deep neural network (DNN) frameworks, enabling the concurrent use of physics as explicit knowledge and data as implicit knowledge, they lack built-in uncertainty quantification, limiting their applicability in scenarios with high noise levels and model discrepancies \cite{luo2025physics}.
Consequently, the application of PINNs in safety-critical, real-world reliability assessment remains underexplored.

\begin{figure*}[t!]
    \centering
    \includegraphics[width=\linewidth]{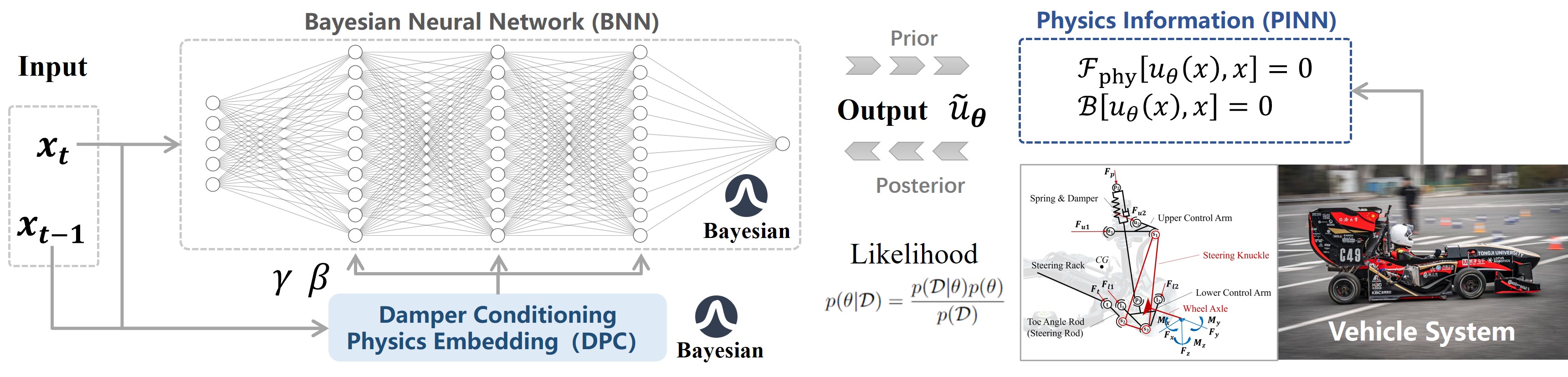}
    \caption{Overview of the proposed \textbf{DBPnet} framework: Sensor data vector $\boldsymbol{x_t}$ is input into the Bayesian neural network (BNN), while $\boldsymbol{x_t}$ and $\boldsymbol{x_{t-1}}$ form the input to the damper characteristic-based physics conditioning (DPC) module together, which embeds physics information into the weight parameters of the BNN.}
    \label{fig:frame}
\end{figure*}

A principled approach to estimate uncertainty in DNNs relies on Bayes' theorem, as exemplified by Bayesian neural networks (BNNs) \cite{kononenko1989bayesian}, which places prior distributions over network weights, allowing posterior inference to capture model uncertainty. 
While traditional Bayesian inference methods require informative prior knowledge about system parameters, they will incur significant computational costs if weakly informative priors are generally applied to the whole range of system parameters \cite{Petra2017}.
Practical implementations therefore rely on efficient approximations such as Monte Carlo dropout \cite{gal2016theoretically}. 
More recently, physics-informed BNNs have been introduced as a hybrid approach that integrates PINNs and Bayesian techniques to quantify uncertainties in both data and models, known as Bayesian physics-informed neural networks (BPINNs) \cite{B-PINNs}.
By employing Hamiltonian Monte Carlos or variational inference (VI) for posterior sampling, BPINNs have demonstrated improved robustness and predictive performance in noisy environments \cite{hou2024improvement}.

Despite these advances, existing methods for vehicle state estimation still face two persistent challenges: \textbf{(1)} non-convergence and instability under significant system noise, and \textbf{(2)} limited estimation accuracy caused by oversimplified physical models. 
To address these gaps, this paper proposes \textbf{DBPnet}, a novel BNN framework inspired by mechanical damper properties, as shown in Figure \ref{fig:frame}. 
Our approach introduces a damper characteristic-based physics conditioning (DPC) embedding mechanism, which leverages the velocity-dependent properties of mechanical dampers to dynamically condition the learning process.
By processing external physical sensor data (e.g., vehicle speed, acceleration, etc.), the DPC module modulates intermediate layers of the main BNN, ensuring that physical priors guide feature learning adaptively rather than through fixed analytical assumptions.
This design improves both estimation accuracy and robustness under diverse operating conditions.
The key contributions of this work are summarized as follows:
\begin{itemize}
    \item \textbf{A Suspension Linkage-Level Modeling Method}, which reconstructs highly nonlinear suspension geometry through refined instantaneous dynamic modeling, improving the reliability of physical information.
    \item \textbf{A Bayesian Physics-Informed Neural Network Framework}, which integrates physical prior knowledge into network training while employing Bayesian inference to mitigate system noise and enhance robustness.
    \item \textbf{A Damping Characteristics-Based Physics Conditioning Embedding Mechanism}, which captures temporal correlations in input signals and dynamically injects them into the network to strengthen physical consistency.
\end{itemize}

The remainder of this paper is organized as follows.
Section \ref{related_works} conducts literature review.
Problem formulation is introduced in Section \ref{setup}.
The proposed \textbf{DBPnet} framework is detailed in Section \ref{method}. 
Experiments and results are presented in Section \ref{experiment}. 
Section \ref{conclusion} concludes the paper and discusses future directions.

\section{Related Work}
\label{related_works}

This section summarizes the existing KF-based methods, PINN-based methods, and physics-informed BNN-based methods in dynamic state estimation.

\subsection{Kalman Filter-Based Methods}

The KF \cite{kalman1960new} is a classical method for state estimation in discrete-time state-space models.
While the original KF assumes linear Gaussian state-space models, some extensions, such as EKF \cite{ribeiro2004kalman}, EnKF \cite{roth2017ensemble}, and unscented Kalman filter (UKF) \cite{wan2000unscented}, have been developed to handle nonlinear and non-Gaussian systems.
With the advances of data-driven state estimation methods, hybrid approaches that combine the theoretical soundness and computational efficiency of KF-based methods with the model-agnostic nature of DNNs have gained increasing attention \cite{zhan2024kfd, bertipaglia2024unscented, hao2025rkfnet, wang2019estimation}.
Building upon KalmanNet \cite{revach2022kalmannet}, several specific extensions have been proposed to provide error covariance \cite{buchnik2023latent, buchnik2024gsp, chen2025maml}.
For instance, Bayesian KalmanNet \cite{dahan2025bayesian} integrated BNNs with KalmanNet and transformed the KF into a stochastic machine learning architecture, employing sampling techniques to predict error covariance without requiring additional domain knowledge.
Despite these advances, KF-based methods often exhibit inconsistency under highly time-varying and strongly nonlinear dynamics.

\subsection{Physics-Informed Neural Network-Based Methods}


PINNs represent a prominent paradigm for incorporating physics knowledge into learning process via physics-based loss functions \cite{wang2025physics}.
Baseline PINN algorithms have achieved remarkable success in solving a wide range of linear and nonlinear PDEs \cite{pinn-energy-ngd}.
To improve robustness and efficiency, numerous new PINN architectures have been proposed, such as conservative PINNs \cite{jagtap2020conservative}, nonlocal PINNs \cite{pang2020npinns}, and adaptive PINNs \cite{Nabian2020}.
In real-world applications, Shi et al. \cite{shi2021physics} constructed physical discrepancy losses based on an existing second-order traffic model to solve traffic state estimation problems.
Xu et al. \cite{xu2022physics} embedded speed and steering models into the loss function to improve generalization under limited training data.
A hybrid PINN-UKF method was proposed by Tan et al. \cite{tan2023vehicle} to mitigate sensor errors and facilitate sensor fusion, leveraging linear time-invariant hypothesis for loss calculation.
EV-PINN \cite{EV-PINN} further demonstrated the use of PINNs for predicting instantaneous battery power and cumulative energy consumption under nonlinear vehicle dynamics.
However, existing PINN-based methods often encounter non-convergence issues and low accuracy due to system uncertainty and nonlinearity.
These limitations present significant challenges for the development and application of PINNs in real-world engineering systems.


\subsection{Physics-Informed Bayesian Neural Network-Based Methods}

Bayesian inference provides a principled framework for learning from noisy and incomplete data \cite{li2014adaptive}.
Motivated by this, Yang et al. \cite{B-PINNs} first combined PINNs with BNNs to address both forward and inverse problems involving linear and nonlinear PDEs under noisy observations.
Since then, BPINNs have demonstrated remarkable performance in uncertainty quantification for physical systems modeled by differential equations \cite{yao2019quality}, inverse problems \cite{EKI-B-PINNs}, and nonlinear dynamical systems \cite{linka2022bayesian}.
For example, in the context of power systems, physics-informed BNNs have demonstrated improved robustness to noise-induced uncertainty, outperforming parsimonious approaches such as sparse identification of nonlinear dynamics \cite{Stock2024}. 
PINNs have also been employed as surrogate models to solve Bayesian inverse problems, significantly reducing computational cost by retaining initial update results to ensure the global nature of the surrogate models \cite{Guan2024}. 
In another work, an offline-online strategy that coupled classical sampling methods with PINN-based approaches for Bayesian inverse problems achieved a substantial reduction in overall computational time while maintaining accuracy \cite{Li2023}. 
Besides, Li et al. \cite{li2025physics} designed a physics-informed BNN framework for forward simulation and unknown parameter inference in non-diffusive thermal modeling using sparse and noisy data.
Nevertheless, existing physics-informed BNN-based models have not yet been thoroughly validated in complex, real-world, safety-critical engineering problems, such as real-time dynamic wheel load estimation.

\section{Problem Formulation}
\label{setup}

This section first introduces the dynamic wheel load estimation problem in this paper. 
Then, we introduce the preliminary to assist the introduction of the proposed \textbf{DBPnet} framework in Section \ref{method}.

\subsection{Vehicle State Estimation}

Vehicle state estimation is a challenging task that can be modeled as an MIMO system \cite{singh2019literature}. 
The complexity arises from the highly nonlinear relationships between sensor inputs and the vehicle's dynamic states \cite{salari2023tire}. 
Vehicle dynamic wheel load is crucial data for vehicle chassis design and automatic control.
Therefore, in this work, we focus on dynamic wheel load estimation, which is formulated as:
\begin{equation}
 [F_{\text{fr}}, F_{\text{fl}}, F_{\text{rr}}, F_{\text{rl}}]^\top = \mathcal{F}_{\text{NN}}(\boldsymbol{x}; \boldsymbol{\theta}),
\end{equation}
where $F_i$ represents the vertical load on each of the four tires, with ``fl" denotes front left, ``fr" denotes front right, ``rr" denotes rear right, and ``rl" denotes rear left. 
The function $\mathcal{F}_{\text{NN}}$ is a neural network with parameters $\boldsymbol{\theta}$ that approximates the true physical relationship. The input vector $\boldsymbol{x}=[\delta, a_{\text{spr}}, a_{\text{unspr}}, d_{\text{sus}}, \dot{d}_{\text{sus}},\boldsymbol{F}_p]^\top$ consists of observable sensor data, including the steering wheel angle $\delta$, sprung mass acceleration $a_{\text{spr}}$, unsprung mass acceleration $a_{\text{unspr}}$, suspension travel distance $d_{\text{sus}}$, suspension velocity $\dot{d}_{\text{sus}}$ and pushrod force $\boldsymbol{F}_p$. 

A key challenge in this MIMO system is the imperfect correspondence between inputs and outputs; for instance, a specific wheel load is primarily influenced by its corresponding suspension inputs, an inductive bias that standard neural networks struggle to learn \cite{zhang2017multi}.

To guide the network's learning process with a physical prior, we utilize a simplified quarter-car model \cite{sieberg2019hybrid}. The dynamic wheel load $F$ is approximated by the following nonlinear function $\mathcal{F}$ \cite{zeng2024analysis}:
\begin{equation}
\label{eq:phy_model}
 F = \mathcal{F}(\boldsymbol{x};\boldsymbol{\lambda}),
\end{equation}
where the input vector $\boldsymbol{\lambda}=[m_{spr},m_{unspr},k_{\{\text{f,r}\}},c_{\{\text{f,r}\}}]^\top$ consists of the vehicle's sprung mass $m_{spr}$, unsprung mass $m_{unspr}$, spring stiffness $k_{\{\text{f,r}\}}$, and damping coefficient $c_{\{\text{f,r}\}}$, respectively.
These parameters often vary between vehicles, and this equation serves as the a prior physical knowledge integrated into our model's loss function.

The estimation of dynamic wheel load presents two fundamental challenges. 
First, accurate estimation is highly dependent on precise chassis modeling, as the wheel–suspension system exhibits strong nonlinearities that are difficult to capture with simplified representations. 
Second, the sensors commonly employed for wheel load estimation are mounted on the chassis and are thus heavily affected by noise and disturbances, which can compromise the reliability of conventional computational methods.

\subsection{Preliminary}
\label{app:pinn}
\subsubsection{Physics Informed Neural Networks}

PINNs are designed to incorporate a prior physical knowledge into neural network training, helping to prevent unreasonable biases and oscillations in the presence of unknown inputs while mitigating network overfitting \cite{TrainingPINNs}. 
Physical information can be integrated into the neural network at various stages, including data preprocessing, the loss function, and backpropagation \cite{Karniadakis2021}. 
Typically, physical laws are expressed as a set of  ordinary differential equations (ODEs), PDEs, equality constraints, or inequality constraints.
Within the setting of this paper, to reach a balance between accuracy and computational speed in practical scenarios, the problem is formulated as a highly nonlinear system of equality equations.
The equations over a domain $\boldsymbol{\Omega} \subset\mathbb{R}^d$ is expressed as \cite{raissi2019physics}:
\begin{equation}
\begin{aligned}
\label{e1}
    \boldsymbol{\mathcal{D}}[\boldsymbol{u}(\boldsymbol{x}),\boldsymbol{x}] = 0, \text{ for }\boldsymbol{x} = (x_1,x_2,\cdots,x_d)^\top \in \boldsymbol{\Omega}, 
\end{aligned}
\end{equation}
where $\boldsymbol{\mathcal{D}}$ is the equality equations of the system and $\boldsymbol{u}$ is modeled as the solution field for all data $\boldsymbol{x} \in \boldsymbol{\Omega}$ satisfying the system constraints.

In PINNs, the solution $\boldsymbol{u}(\boldsymbol{x})$ is approximated by a deep neural network, denoted as $\boldsymbol{\hat{u}}(\boldsymbol{x}; \boldsymbol{\theta})$, where $\boldsymbol{\theta}$ represents the set of all learnable parameters.

The discrepancy between the results calculated by the neural network and by physical model is represented by a residual $\boldsymbol{\mathcal{R}}(\boldsymbol{x}; \boldsymbol{\theta})$, which is defined by:
\begin{equation}
\label{e2}
    \boldsymbol{\mathcal{R}}(\boldsymbol{x}; \boldsymbol{\theta}):=\boldsymbol{\mathcal{D}}[\boldsymbol{\hat{u}}(\boldsymbol{x}; \boldsymbol{\theta}), \boldsymbol{x}].
\end{equation}

The parameters $\boldsymbol{\theta}$ are optimized by minimizing a physical loss $\mathcal{L}_{\text{p}}(\boldsymbol{\theta})$:
\begin{equation}
\begin{aligned}
\label{e3}
    \mathcal{L}_{\text{p}}(\boldsymbol{\theta}) =  
    \frac{1}{N_f} \sum_{i=1}^{N_f} \| \boldsymbol{\mathcal{R}}(\boldsymbol{x}_f^i; \boldsymbol{\theta}) \|_2^2,
\end{aligned}
\end{equation}
where $\{ \boldsymbol{x}_f^i \}_{i=1}^{N_f}$ is a set of $N_f$ collocation points sampled from the domain $\boldsymbol{\Omega}$.

A Data-driven loss $\mathcal{L}_{\text{d}}(\boldsymbol{\theta})$ is presented as:
\begin{equation}
\label{e6}
    \mathcal{L}_{\text{d}}(\boldsymbol{\theta}) = \frac{1}{N_d} \sum_{i=1}^{N_d} \| \boldsymbol{\hat{u}}(\boldsymbol{x}_d^i; \boldsymbol{\theta}) - \boldsymbol{u}^i \|_2^2,
\end{equation}
where $\boldsymbol{u}^i$ here is the data ground truth.
Total loss $\mathcal{L}(\boldsymbol{\theta})$ becomes weighted sum of $\mathcal{L}_{\text{p}}(\boldsymbol{\theta})$ and $\mathcal{L}_{\text{d}}(\boldsymbol{\theta})$.
It should be noted that in conventional PINNs for solving PDEs, the physical loss also includes boundary conditions. However, in this paper, the problem is modeled as a system of equality equations, where boundary conditions have been embedded into the system equations, so no additional boundary constraints are required.

\subsubsection{Revolute-Sphere-Sphere-Revolute Structure Model}
\label{pre:RSSR}

\begin{figure}[t!]
    \centering
    \includegraphics[width=0.88\linewidth]{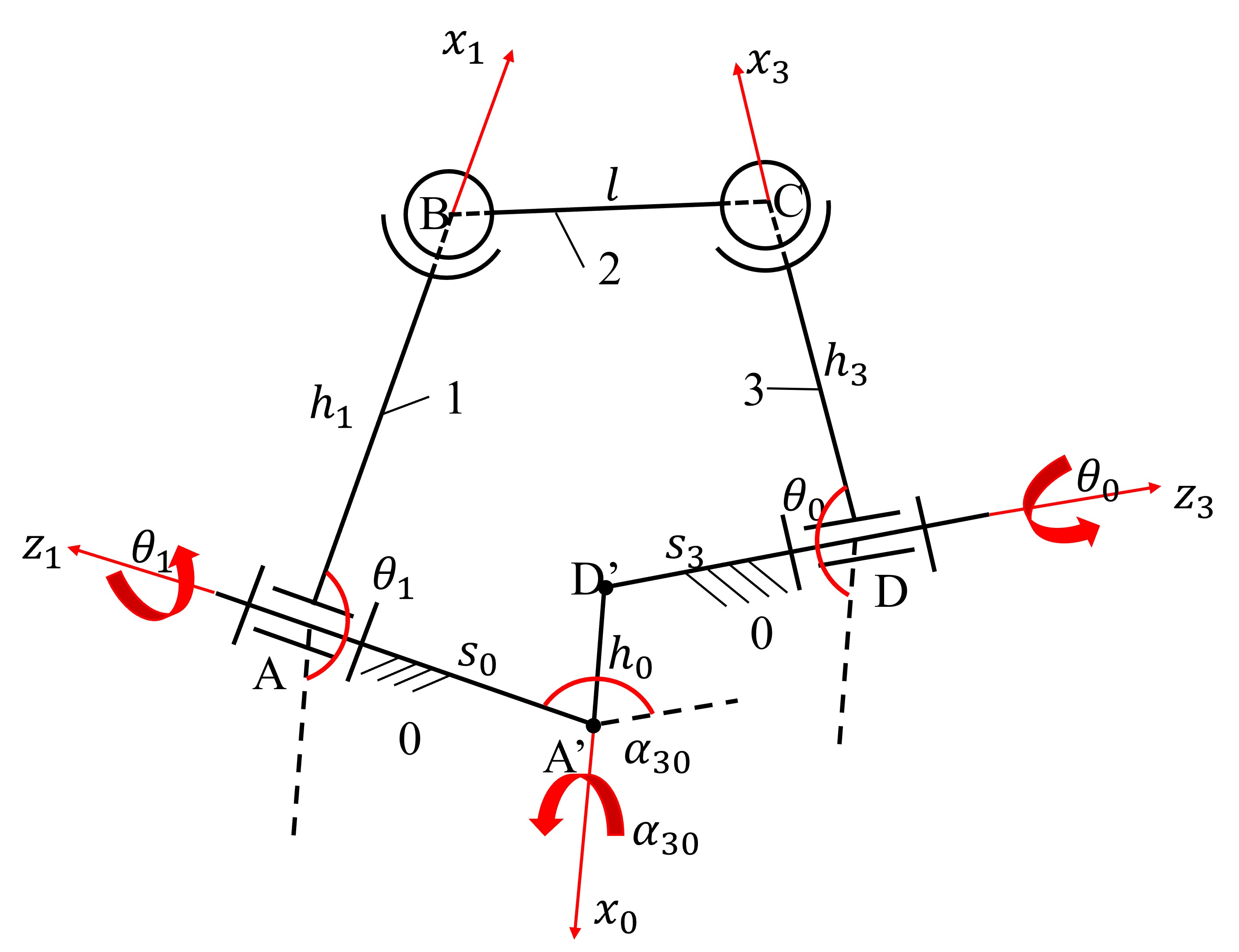}
    \caption{Revolute-sphere-sphere-revolute (RSSR) structure model.}
    \label{fig:RSSR}
\end{figure}

The spatial revolute-sphere-sphere-revolute (RSSR) model is a classic parallel mechanism, as illustrated in Figure \ref{fig:RSSR}, consisting of two revolute joints and two spherical joints \cite{zeng2023wheel}.
In the figure, $A$, $B$, $C$, and $D$ denote four hinges, while $0$, $1$, $2$, and $3$ represent base frame and three links. The parameters $h_1$, $h_3$, and $l$ stand for the lengths of the links. $A'D'$ is the common perpendicular between the two axes $z_1$ and $z_3$. The parameter $h_0$ denotes the distance between axes $z_1$ and $z_3$. The parameters $s_0$ and $s_3$ represent the distances from the hinge points to $A'D'$. The angles $\theta_1$ and $\theta_0$ are the included angles between links 1, 3 and the base frame, respectively. $\alpha_{30}$ denotes the included angle between the $z_1$ and $z_3$ axes.
The coordinates of points $B$ and $C$ can be expressed as:
\begin{equation}
\begin{pmatrix}
x_B \\
y_B \\
z_B
\end{pmatrix}
=
\begin{pmatrix}
0 \\
0 \\
s_0
\end{pmatrix}
+ [C_{01}]
\begin{pmatrix}
h_1 \\
0 \\
0
\end{pmatrix}
=
\begin{pmatrix}
h_1 \cos\theta_1 \\
h_1 \sin\theta_1 \\
s_0
\end{pmatrix}
,
\end{equation}

\begin{equation}
\begin{aligned}
\begin{pmatrix}
x_C \\
y_C \\
z_C
\end{pmatrix}
& =
\begin{pmatrix}
-h_0 \\
0 \\
0
\end{pmatrix}
+ [C_{30}]^{-1}
\begin{pmatrix}
h_3 \\
0 \\
s_3
\end{pmatrix} \\
& =
\begin{pmatrix}
-h_0 + h_3 \cos\theta_0 \\
-h_3 \cos\alpha_{30} \sin\theta_0 + s_3 \sin\alpha_{30} \\
h_3 \sin\alpha_{30} \sin\theta_0 + s_3 \cos\alpha_{30}
\end{pmatrix}
,
\end{aligned}
\end{equation}
where $[C_{ij}]$ means the direction cosine matrix from coordinate $i$ to $j$, which can be expressed as:
\begin{equation}
\begin{aligned}
C_{ij}^{(\theta, \alpha)} 
&= 
\begin{bmatrix}
\cos\theta & -\sin\theta & 0 \\
\sin\theta & \cos\theta & 0 \\
0 & 0 & 1
\end{bmatrix}
\begin{bmatrix}
1 & 0 & 0 \\
0 & \cos\alpha & -\sin\alpha \\
0 & \sin\alpha & \cos\alpha
\end{bmatrix} \\
&= 
\begin{bmatrix}
\cos\theta & -\sin\theta\cos\alpha & \sin\theta\sin\alpha \\
\sin\theta & \cos\theta\cos\alpha & -\cos\theta\sin\alpha \\
0 & \sin\alpha & \cos\alpha
\end{bmatrix}
,
\end{aligned}
\end{equation}
where $\theta$ and $\alpha$ are rotation angles on the two coordinates.
Given the angle $\theta_1$ of the mechanism, the angle $\theta_0$ can be solved by the following equation:
\begin{equation}
A \sin\theta_0 + B \cos\theta_0 + C = 0,
\end{equation}
where
\begin{equation}
\begin{aligned}
A &= \cos\alpha_{30} \sin\theta_1 - \frac{s_0 \sin\alpha_{30}}{h_1}, \\
B &= -\left( \frac{h_0}{h_1} + \cos\theta_1 \right), \\
C &= \frac{h_1^2 - l^2 + h_3^2 + h_0^2 + s_3^2 + s_0^2 - 2 s_s s_0 \cos\alpha_{30}}{2 h_1 h_3} \\
    &+ \frac{h_0 \cos\theta_1 - s_3 \sin\alpha_{30} \sin\theta_1}{h_3}.
\end{aligned}
\end{equation}
Replacing one revolute joint with a prismatic joint and setting the link length to approach zero enables calculation using the above equations.

\section{Methodology}
\label{method}

This section first builds a refined suspension linkage-level modeling (SLLM) that reconstructs the nonlinear geometric and mechanical relationships within the suspension system, which captures the instantaneous dynamic behavior of the suspension with higher fidelity, providing a reliable physical basis for dynamic wheel load estimation. 
Then, we introduce the proposed \textbf{DBPnet} framework as shown in Figure \ref{fig:frame}. 
We elaborate on the foundational BPINN framework, with a specific focus on VI for posterior estimation, and a novel conditioning module inspired by the dynamic properties of mechanical dampers.

\begin{figure}[ht]
    \centering
    \includegraphics[width=1\linewidth]{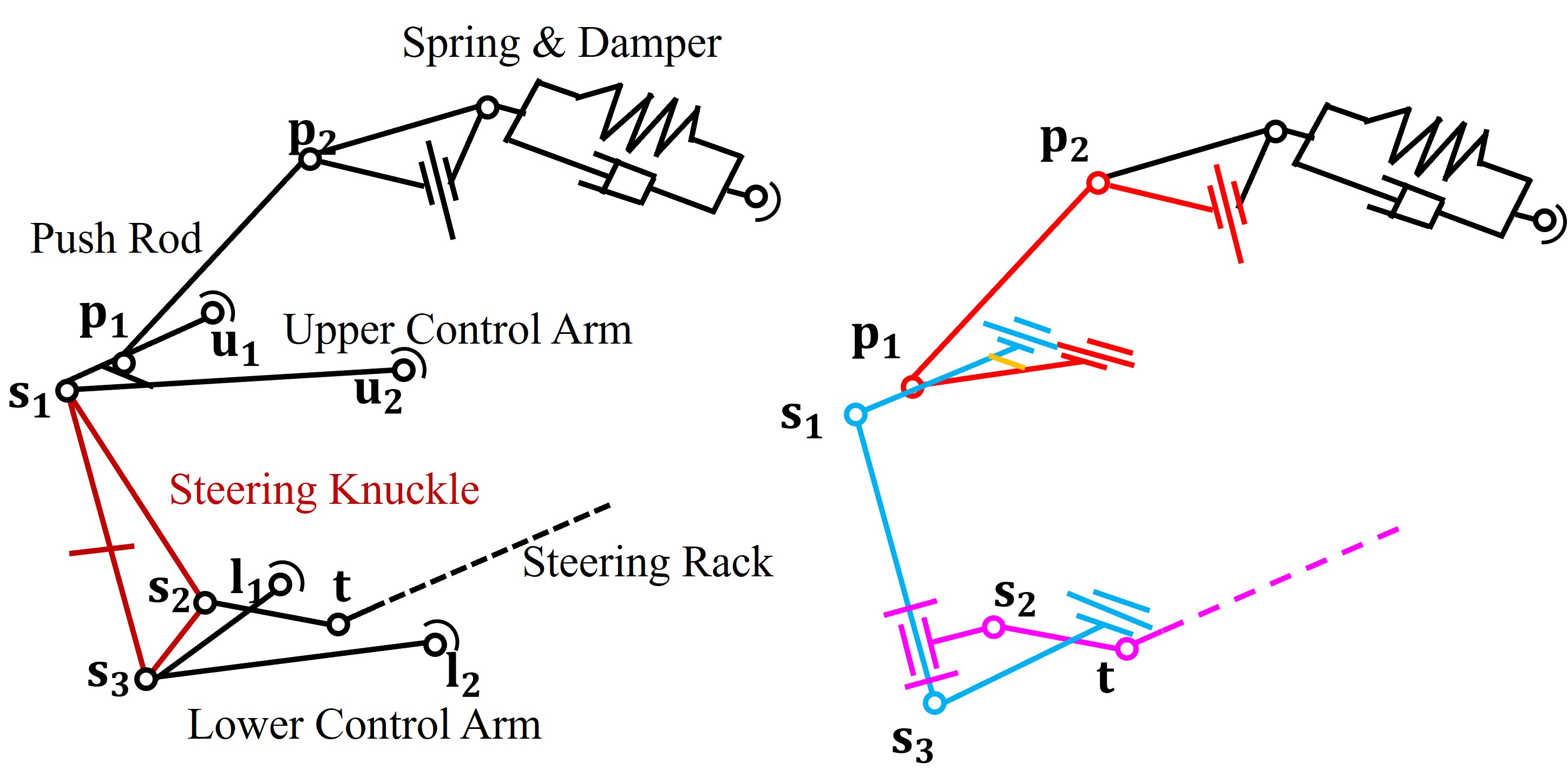}
    \caption{Front suspension model for wheel load estimation: Two-degree-of-freedom (2 DOF) with bumping and steering. The hinge with a semicircular arc shown in the figure is fixed to the vehicle body.}
    \label{fig:sus}
\end{figure}

\subsection{Suspension Linkage-Level Modeling}

Tire dynamic load refers to the dynamic vertical force exerted on the tire during vehicle motion. 
Due to the complexity of the vehicle chassis structure (incorporating components such as springs and dampers), directly calculating this load based solely on vehicle body attitude and acceleration introduces significant errors. 
This approach fails to accurately reflect the actual force experienced by the tire, thereby limiting the ability to properly evaluate chassis performance.

During vehicle operation, the spring \& damper assembly undergoes expansion and contraction in response to relative displacement between the road surface and vehicle body. 
Since vehicle suspensions typically consist of multiple connecting rods forming complex single-degree-of-freedom (1 DOF) or two-degree-of-freedom (2 DOF) structures, their geometric parameters evolve dynamically during driving \cite{zeng2023dynamic}. 
The hinge joints of the suspension system, are commonly referred to as ``hard points''. 
Taking a double-wishbone suspension as an example, we simplify it to the linkage structure shown in Figure \ref{fig:sus}, on the left is the original structural schematic of the double wishbone suspension. 
In this study, the hard points are defined as follows:
$\bold{u_1}$ and $\bold{u_2}$ (front and rear points of the upper control arm, UCA);
$\bold{l_1}$ and $\bold{l_2}$ (front and rear points of the lower control arm, LCA);
$\bold{p_1}$ and $\bold{p_2}$ (lower and upper joints of the spring-damper assembly);
$\bold{t}$ (hinge joint between the steering knuckle and steering rack);
and $\bold{s_1}$, $\bold{s_2}$, and $\bold{s_3}$ (upper, front, and lower points of the steering knuckle).
We simplify it using a RSSR model for the rapid solution of the dynamic geometric structure, which is shown on the right.
The geometry of the above structure can be directly determined by the compression of the spring \& damper and the extension of the steering rack (i.e., the steering angle). The three distinct colors in the figure represent three parallel submechanisms, which is more convenient and efficient than solving for the positions of the original spatial points one by one.

\begin{figure}[ht!]
    \centering
    \includegraphics[width=0.55\linewidth]{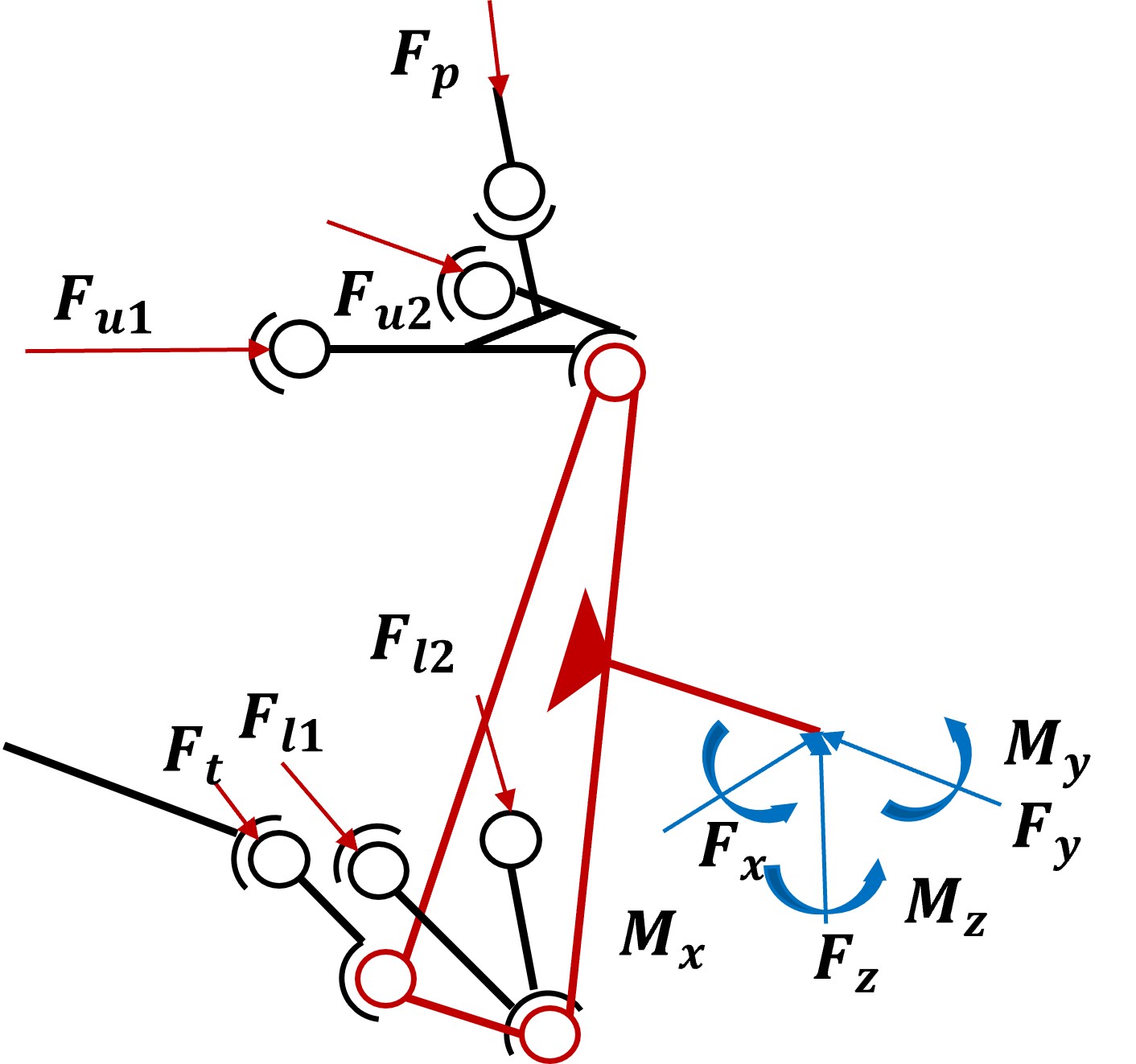}
    \caption{Force diagram of the double wishbone suspension. Red: forces in suspension links. Blue: external forces and torques transmitted from the tire to the suspension via the wheel axle.}
    \label{fig:force}
\end{figure}

Force diagram is presented in Figure \ref{fig:force}.
For computational simplicity, the mass of individual suspension links is neglected, and only the unsprung mass $m_u$, which includes the wheel and steering knuckle, is considered.
We denote $CG$ as the center of mass; $\boldsymbol{F_p}$, $\boldsymbol{F_{u1}}$, $\boldsymbol{F_{u2}}$, $\boldsymbol{F_t}$, $\boldsymbol{F_{l1}}$, $\boldsymbol{F_{l2}}$ represent suspension force vectors; $\boldsymbol{F_x}$, $\boldsymbol{F_y}$, $\boldsymbol{F_z}$ are tire forces transmitted from the wheel; and $\boldsymbol{M_x}$, $\boldsymbol{M_y}$, $\boldsymbol{M_z}$ are moments about the wheel axle. 

We define the vector from $CG$ to hard point $i$ as $\overrightarrow{i}$ (e.g., $\overrightarrow{\bold{u_1}}$) and the direction vector from hard point $i$ to $j$ as $\overrightarrow{ij}$ (e.g., $\overrightarrow{\bold{u_1}\bold{s_1}}$). 
Force vectors can thus be expressed as:
\begin{equation}
\begin{aligned}
\label{eq2}
    &\boldsymbol{F}_p = F_p \cdot \overrightarrow{\bold{p_2}\bold{p_1}}, \quad 
    \boldsymbol{F}_{u1} = F_{u1} \cdot \overrightarrow{\bold{u_1}\bold{s_1}}, \\
    &\boldsymbol{F}_{u2} = F_{u2} \cdot \overrightarrow{\bold{u_2}\bold{s_1}}, \quad 
    \boldsymbol{F}_{t} = F_{t} \cdot \overrightarrow{\bold{t}\bold{s_2}}, \\
    &\boldsymbol{F}_{l_1} = F_{l_1} \cdot \overrightarrow{\bold{l_1}\bold{s_3}}, \quad 
    \boldsymbol{F}_{l_2} = F_{l_2} \cdot \overrightarrow{\bold{l_2}\bold{s_3}},
\end{aligned}
\end{equation}
where unbolded coefficients represent scalar magnitudes of the corresponding forces. 

The suspension force matrix is:
\begin{equation}
    \boldsymbol{F_{\text{sus}}} = \left[ \boldsymbol{F}_p, \boldsymbol{F}_{u_1}, \boldsymbol{F}_{u_2}, \boldsymbol{F}_{t}, \boldsymbol{F}_{l_1}, \boldsymbol{F}_{l_2} \right].
\end{equation}

Moments of these forces about $CG$ are calculated as:
\begin{equation}
\begin{aligned}
    &\boldsymbol{M}_p = \overrightarrow{\bold{p_2}} \times \boldsymbol{F_p}, \quad 
    \boldsymbol{M}_{u_1} = \overrightarrow{\bold{u_1}} \times \boldsymbol{F_{u_1}}, \\
    &\boldsymbol{M}_{u_2} = \overrightarrow{\bold{u_2}} \times \boldsymbol{F_{u_2}}, \quad 
    \boldsymbol{M}_{t} = \overrightarrow{\bold{t}} \times \boldsymbol{F_t}, \\
    &\boldsymbol{M}_{l_1} = \overrightarrow{\bold{l_1}} \times \boldsymbol{F_{l_1}}, \quad 
    \boldsymbol{M}_{l_2} = \overrightarrow{\bold{l_2}} \times \boldsymbol{F_{l_2}}.
\end{aligned}
\end{equation}

The moment matrix is:
\begin{equation}
    \boldsymbol{M_{\text{sus}}} = \left[ \boldsymbol{M}_p, \boldsymbol{M}_{u_1}, \boldsymbol{M}_{u_2}, \boldsymbol{M}_{t}, \boldsymbol{M}_{l_1}, \boldsymbol{M}_{l_2} \right].
\end{equation}

Force and moment equilibrium relationships yield:
\begin{equation}
\label{eq24}
    \sum_{i \in \{p,u_1,u_2,t,l_1,l_2\}} \boldsymbol{F}_i +m_u \boldsymbol{a}_u = \sum_{j \in \{x,y,z\}} \boldsymbol{F}_j,
\end{equation}
\begin{equation}
\label{eq25}
    \sum_{i \in \{p,u_1,u_2,t,l_1,l_2\}} \boldsymbol{M}_i+I_u \boldsymbol{\beta}_u = \sum_{j \in \{x,y,z\}} \boldsymbol{M}_j,
\end{equation}
where $m_u$ and $I_u$ are mass and moment of inertia about $CG$ of the unsprung mass, $\boldsymbol{a}_u$ and $\boldsymbol{\beta}_u$ are linear and angular acceleration of the unsprung mass.

The front suspension has 2 DOF: steering and bumping. 
We denote the steering rack displacement as $x_a$ and the expansion/contraction of the spring as $x_d$, both of which are measurable by sensors. 
These parameters directly influence suspension geometry, so hard point coordinates are expressed as functions of $x_a$ and $x_d$ (e.g., $\overrightarrow{\bold{u_1}(x_a,x_d)}$, $\overrightarrow{\bold{u_2}(x_a,x_d)}$), which are solved by RSSR models mentioned in Section \ref{pre:RSSR}.

By installing sensors on the pushrod, we obtain $\boldsymbol{F}_p$. 
Since the internal stress of the suspension structure is difficult to measure accurately using spring-damper calculations, we directly incorporate data from a pushrod force sensor as input to reduce error.
Furthermore, tire forces $\boldsymbol{F}_x$ and $\boldsymbol{F}_y$ are related to $\boldsymbol{F}_z$ according to the magic formula \cite{pacejka2005tire}. 
Therefore, $\boldsymbol{F}_z$ is solved using $x_a$, $x_d$, $\dot{x_d}$, and $\boldsymbol{a}_u$ as:
\begin{equation}
\boldsymbol{F}_z=f(\boldsymbol{x})=f([x_a,x_d,\dot{x}_d,\boldsymbol{a}_u,\boldsymbol{F}_p]^\top),
\end{equation}
where $f$ represents all the solving processes from Equation (\ref{eq2}) to Equation (\ref{eq25}).
Then, the complex suspension dynamics will be integrated into the BNN as physical guidance.

\subsection{Bayesian Physics-Informed Neural Network}

\subsubsection{Physics-Informed Bayesian Framework}

Our framework is built upon the foundation of BPINNs, which synergizes the uncertainty quantification capabilities of BNNs with the physical consistency enforced by PINNs.

A standard PINN embeds physical laws into its loss function to constrain the solution space \cite{Karniadakis2021}. 
We use a general form for physical constraints:
\begin{equation}
\label{e7}
    \mathcal{F}_{\text{phy}}[u_{\theta}(\boldsymbol{x}); \lambda] = 0, \quad \boldsymbol{x} \in \Omega, \\
\end{equation}
and enforce consistency with the available observations $\mathcal{D}$:
\begin{equation}
    \mathcal{B}[u_{\theta}(\boldsymbol{x}); \lambda] = 0, \quad \text{for each data pair } (\boldsymbol{x}, u) \in \mathcal{D},
\end{equation}
where $u_{\theta}(\boldsymbol{x})$ is the neural network solution with parameters $\theta$, which approximates the true physical relationship, $\lambda$ is the vector of parameters in the physics law, the input $\boldsymbol{x}$ belongs to the operational domain $\Omega \subseteq \mathbb{R}^d$, $d$ is the dimension of $\boldsymbol{x}$, $\mathcal{F}_{\text{phy}}$ is the physics-based residual operator derived from the system's governing equations, and $\mathcal{B}$ is the constraint operator that penalizes mismatches with observations.

The loss function is a weighted sum of a data-driven term and a physics-based term:
\begin{equation}
 \mathcal{L}_{\text{PINN}} = w_{\text{data}} \mathcal{L}_{\text{data}} + w_{\text{phy}}\mathcal{L}_{\text{phy}},
\end{equation}
where $\mathcal{L}_{\text{data}}$ measures the discrepancy between network predictions and training data, and $\mathcal{L}_{\text{phy}}$ penalizes violations of the known physical laws \cite{TrainingPINNs}.

To account for system noise and model uncertainty, we adopt a Bayesian approach, which learns a posterior distribution $p(\theta | \mathcal{D})$ over the weights given the observed dataset $\mathcal{D}$ instead of learning point estimates for the network weights $\theta$ \cite{mackay1995bayesian}.
According to Bayes' theorem, the posterior probability of the parameters $\theta$ given the data $\mathcal{D}$ is expressed as:
\begin{equation}
 p(\theta | \mathcal{D}) = \frac{p(\mathcal{D} | \theta) p(\theta)}{p(\mathcal{D})},
\end{equation}
where $p(\theta)$ is the prior distribution of the network weights, representing our initial beliefs about the parameters before observing any data, $p(\mathcal{D} | \theta)$ is the likelihood function, which quantifies the probability of observing the data $\mathcal{D}$ given a specific set of parameters $\theta$, and $p(\mathcal{D})$ is the evidence (or marginal likelihood), a normalizing constant that ensures the posterior distribution integrates to one.

\subsubsection{Variational Inference with Physics-Informed Likelihood}

In the context of BPINNs, the observed dataset $\mathcal{D}$ consists of noisy measurements of the solution $u$, the forcing term $f$, and the boundary conditions $b$. Specifically, as described in \cite{B-PINNs}, $\mathcal{D}$ is the union of these measurement sets as:
\begin{equation}
\mathcal{D}=\mathcal{D}_{u}\cup\mathcal{D}_{f}\cup\mathcal{D}_{b},
\end{equation}
where $\mathcal{D}_{u}=\{(x_{u}^{(i)},\tilde{u}^{(i)})\}_{i=1}^{N_{u}}$, $\mathcal{D}_{f}=\{(x_{f}^{(i)},\tilde{f}^{(i)})\}_{i=1}^{N_{f}}$, and $\mathcal{D}_{b}=\{(x_{b}^{(i)},\tilde{b}^{(i)})\}_{i=1}^{N_{b}}$ represent the noisy observations of $u$, $f$, and $b$ at respective spatial locations. 

We assume these measurements are corrupted by independent Gaussian noise, for example:
\begin{equation}
\tilde{u}^{(i)}=u(x_{u}^{(i)})+\epsilon_{u}^{(i)},
\end{equation}
where $\epsilon_{u}^{(i)}$ is Gaussian noise with zero mean and known standard deviation $\sigma_{u}^{(i)}$. Similar formulations apply for $\tilde{f}^{(i)}$ and $\tilde{b}^{(i)}$.

The likelihood $p(\mathcal{D} | \theta)$ is then formed by combining the likelihoods from each observation type, assuming independence:
\begin{equation}
P(\mathcal{D}|\theta)=P(\mathcal{D}_{u}|\theta)P(\mathcal{D}_{f}|\theta)P(\mathcal{D}_{b}|\theta).
\end{equation}

Each component of the likelihood is a product of Gaussian probability density functions, reflecting the noisy measurements:
\begin{align}
P(\mathcal{D}_{u}|\theta)&=\prod_{i=1}^{N_{u}}\frac{1}{\sqrt{2\pi{\sigma_{u}^{(i)}}^{2}}}\exp\left(-\frac{(\tilde{u}(x_{u}^{(i)};\theta)-\tilde{u}^{(i)})^{2}}{2{\sigma_{u}^{(i)}}^{2}}\right), \\
P(\mathcal{D}_{f}|\theta)&=\prod_{i=1}^{N_{f}}\frac{1}{\sqrt{2\pi{\sigma_{f}^{(i)}}^{2}}}\exp\left(-\frac{(\tilde{f}(x_{f}^{(i)};\theta)-\tilde{f}^{(i)})^{2}}{2{\sigma_{f}^{(i)}}^{2}}\right), \\
P(\mathcal{D}_{b}|\theta)&=\prod_{i=1}^{N_{b}}\frac{1}{\sqrt{2\pi{\sigma_{b}^{(i)}}^{2}}}\exp\left(-\frac{(\tilde{b}(x_{b}^{(i)};\theta)-\tilde{b}^{(i)})^{2}}{2{\sigma_{b}^{(i)}}^{2}}\right),
\end{align}
where $\tilde{u}(x;\theta)$, $\tilde{f}(x;\theta)=\mathcal{F}_{\text{phy}}(\tilde{u}(x;\theta);\lambda)$, and $\tilde{b}(x;\theta)=\mathcal{B}(\tilde{u}(x;\theta);\lambda)$ are the predictions from the neural network and the physics-informed operators.

The unnormalized posterior distribution is then proportional to the product of the likelihood and the prior:
\begin{equation}
P(\theta|\mathcal{D}) \propto P(\mathcal{D}|\theta)P(\theta).
\end{equation}

Since the true posterior $p(\theta | \mathcal{D})$ is often intractable, we employ VI to approximate it, which introduces a simpler, tractable distribution $Q(\theta;\zeta)$, parameterized by $\zeta$, and minimizes the Kullback-Leibler (KL) divergence between this approximate posterior and the true posterior:
\begin{equation}
\min_{Q(\theta;\zeta)} \quad\text{KL}\left(Q(\theta;\zeta) \, || \, P(\theta | \mathcal{D})\right).
\end{equation}

Minimizing this KL divergence is equivalent to maximizing the evidence lower bound (ELBO) \cite{blundell2015weight}, which results in the following optimization objective for the network parameters $\zeta$:
\begin{equation}
\max_{Q(\theta;\zeta)}\quad \mathbb{E}_{Q(\theta;\zeta)} \left[ \log P(\mathcal{D}|\theta) \right] - \text{KL}(Q(\theta;\zeta) \parallel P(\theta)),
\end{equation}
where, the first term, the expected log-likelihood, encourages the approximate posterior to explain the observed data, aligning with the data-fitting loss terms; and the second term acts as a regularizer, penalizing divergence from the prior distribution of the weights.

In practice, a common choice for the approximate posterior $Q(\theta;\zeta)$ in deep learning is the mean-field Gaussian approximation \cite{B-PINNs}, where each component of $\theta$ is assumed to be an independent Gaussian distribution:
\begin{equation}
Q(\theta;\zeta)=\prod_{i=1}^{d_{\theta}}q(\theta_{i};\zeta_{\mu,i},\zeta_{\rho,i}),
\label{eq:mean_field_gaussian}
\end{equation}
where $d_{\theta}$ is the dimension of $\theta$, $\zeta=(\zeta_{\mu},\zeta_{\rho})$, and $q(\theta_{i};\zeta_{\mu,i},\zeta_{\rho,i})$ is the density of a one-dimensional Gaussian distribution with mean $\zeta_{\mu,i}$ and standard deviation $\ln(1+\exp(\zeta_{\rho,i}))$. The latter parametrization for the standard deviation ensures it is always positive.

Then, the objective function for VI is given as \cite{B-PINNs}:
\begin{equation}
\begin{aligned}
&D_{\text{KL}}(Q(\theta;\zeta)||P(\theta|\mathcal{D})) \approx\\ &\mathbb{E}_{\theta \sim Q(\theta;\zeta)} \left[ \ln Q(\theta;\zeta) - \ln P(\theta) - \ln P(\mathcal{D}|\theta) \right].
\label{eq:vi_loss}
\end{aligned}
\end{equation}

This expectation is then approximated using Monte Carlo sampling. After training, samples from $Q(\theta;\zeta)$ are used to generate predictions and quantify uncertainty of $\tilde{u}(x)$.

\subsubsection{Normal-Sigmoid-Dropout Method}

Recent works \cite{Zhang2019} have shown that training a neural network with dropout is equivalent to a form of VI. 
However, standard dropout, where neurons are randomly set to zero, can lead to poor convergence in complex systems. 
To address this, we propose a normal-sigmoid-dropout (NS-Dropout) method. 
For a given layer's neuron activations $x_{n+1}$, we introduce a stochastic noise matrix $H$. 
Its elements $h_i$ are sampled from a normal distribution, $h_i \sim \mathcal{N}(0, \sigma^2)$, where $\sigma$ is a tunable hyperparameter. 
The noise matrix is:
\begin{equation}
 H = \frac{1}{2}\phi(H) + \frac{\boldsymbol{I}}{2},
\end{equation}
where $\phi$ is the sigmoid function and $\boldsymbol{I}$ is an identity matrix. 
This transformation bounds the multiplicative noise for each neuron within the range $(\frac{1}{2}, 1)$, preventing the complete deactivation of neurons. 
The dropout-applied neuron values $x'_{n+1}$ are then computed as:
\begin{equation}
 x'_{n+1} = H \odot x_{n+1},
\end{equation}
where $\odot$ denotes the element-wise product. 
This method provides a more stable way to inject stochasticity, facilitating robust VI.

\subsection{Damper-Characteristic Physics Conditioning Embedding}

\begin{figure}[t!]
    \centering
    \includegraphics[width=0.6\linewidth]{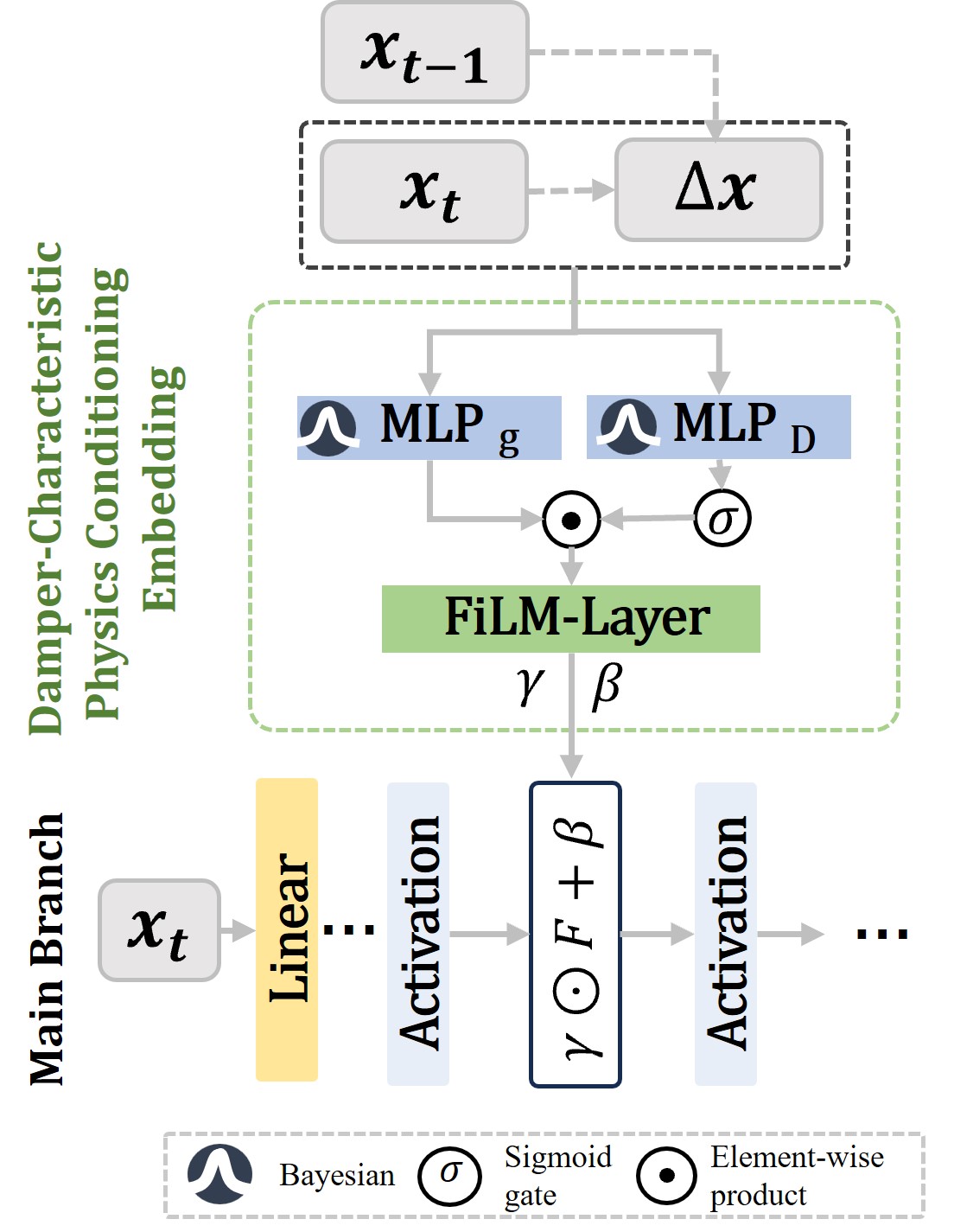}
    \caption{Structure of the proposed damper characteristic-based physics conditioning (DPC) module. }
    \label{fig:damper}
\end{figure}

To deeply embed the physical dynamics into our model, we propose a novel physics-informed conditioning encoder, termed the DPC module, which is inspired by the behavior of a mechanical damper and its purpose is to generate dynamic conditioning parameters for the main BNN, which is shown in Figure \ref{fig:damper}.
The proposed DPC module leverages the FiLM technique \cite{perez2018film}, which applies a feature-wise affine transformation $\mathcal{G}(F) = \gamma \odot F + \beta$ to the intermediate activation Vector $F$ of a network.

The DPC encoder is designed to produce these modulation parameters, i.e. $\gamma$ and $\beta$, from the vehicle's real-time state dynamics. The encoder takes a time-series of sensor inputs, specifically the current state $\boldsymbol{x}_t$ and the previous state $\boldsymbol{x}_{t-1}$ as its input. Its operation begins by explicitly modeling the velocity-dependent force of damper. It estimates the discrete-time ``velocity" of the state by calculating the difference between consecutive time steps as $\Delta \boldsymbol{x}_t = \boldsymbol{x}_t - \boldsymbol{x}_{t-1}$.

Following this, the encoder utilizes two parallel pathways to learn a sophisticated, state-dependent damping representation. The first pathway learns a nonlinear $D_t$ by processing both the current state and its rate of change through a multi-layer perceptron (MLP):
\begin{equation}
    D_t = \text{MLP}_D([\Delta \boldsymbol{x}_t, \boldsymbol{x}_t]).
\end{equation}

Simultaneously, the second pathway processes only the current state $\boldsymbol{x}_t$ to generate a gating signal $g_t$, which is constrained between 0 and 1 by a sigmoid function $\sigma$:
\begin{equation}
    g_t = \sigma(\text{MLP}_g(\boldsymbol{x}_t)).
\end{equation}

This gate acts as a dynamic control valve, allowing the model to learn how much of the damping effect to apply based on the vehicle's current state.
The final output is:
\begin{equation}
    [\gamma_t, \beta_t] = \text{Linear}(g_t \odot D_t),
\end{equation}
where $\odot$ is element-wise product.

These generated parameters $\gamma_t$ and $\beta_t$ are then injected into the main \textbf{DBPnet}. 
The BPINN in this paper is constructed with MLP, and at each layer, the activation vectors are modulated by the parameters produced by the DPC encoder. This allows the physical dynamics encoded by the DPC module to directly influence and guide the feature extraction process throughout the entire network. 

\subsection{Algorithm}

Algorithm \ref{alg:dbpnet} presents the proposed \textbf{DBPnet} framework. 
The implementation details of each module have been explained.

\begin{algorithm}[t!]
\caption{DBPnet: Damper Characteristics-Based Bayesian Physics-Informed Neural Network}
\label{alg:dbpnet}
\begin{algorithmic}[1]
\small
\REQUIRE Dataset $\mathcal{D} = \{(\bm{x}_t, \bm{x}_{t-1}, \bm{u})\}$ where $\bm{x}_t = [\delta, a_{\text{spr}}, a_{\text{unspr}}, d_{\text{sus}}, \dot{d}_{\text{sus}}, F_p]^\top$, physical parameters $\bm{\lambda}$, hyperparameters $L, K, S, \eta, \sigma_n, w_d, w_p$
\ENSURE Trained parameters $\bm{\zeta}^*, \bm{\Theta}_{\text{DPC}}^*$
 
\Statex \textbf{Phase I: Suspension Linkage-Level Modeling}
\STATE Solve RSSR closure equations to obtain hard point positions as functions of $(x_a, x_d)$
\STATE Derive physics residual $\mathcal{F}_{\text{phy}}(\tilde{\bm{u}}, \bm{x}; \bm{\lambda})$ from force/moment equilibrium
 
\Statex \textbf{Phase II: Training}
\STATE Initialize variational parameters $\bm{\zeta} = (\bm{\zeta}_\mu, \bm{\zeta}_\rho)$ and DPC parameters $\bm{\Theta}_{\text{DPC}}$
\FOR{each epoch, each mini-batch $\mathcal{B} \subset \mathcal{D}$}
    \FOR{$k = 1$ to $K$}
        \STATE $\mathcal{L}_d \gets 0$;\; $\mathcal{L}_p \gets 0$
        \STATE Sample BNN weights: $\theta_i \gets \zeta_{\mu,i} + \ln(1{+}\exp(\zeta_{\rho,i})) \cdot \varepsilon_i$, \; $\varepsilon_i \sim \mathcal{N}(0,1)$
        \FOR{each $(\bm{x}_t, \bm{x}_{t-1}, \bm{u}) \in \mathcal{B}$}
            \STATE \textit{// DPC module}
            \STATE $\bm{D}_t \gets \mathrm{MLP}_D([\bm{x}_t {-} \bm{x}_{t-1},\; \bm{x}_t])$; \quad $\bm{g}_t \gets \sigma(\mathrm{MLP}_g(\bm{x}_t))$
            \STATE $[\bm{\gamma}_t, \bm{\beta}_t] \gets \mathrm{Linear}(\bm{g}_t \odot \bm{D}_t)$
            \STATE \textit{// Conditioned BNN forward pass}
            \STATE $\bm{h} \gets \bm{x}_t$
            \FOR{$l = 1$ to $L$}
                \STATE $\bm{h} \gets (\bm{\gamma}_t \odot \mathrm{Act}(\bm{W}_l \bm{h} {+} \bm{b}_l) + \bm{\beta}_t) \odot \bm{H}$, \; $\bm{H} {=} \tfrac{1}{2}\sigma(\bm{\epsilon}){+}\tfrac{1}{2}\bm{I}$, \; $\bm{\epsilon} {\sim} \mathcal{N}(\bm{0}, \sigma_n^2\bm{I})$
            \ENDFOR
            \STATE $\tilde{\bm{u}} \gets \bm{W}_{\mathrm{out}} \bm{h} + \bm{b}_{\mathrm{out}}$
            \STATE $\mathcal{L}_d \mathrel{+}= \|\tilde{\bm{u}} {-} \bm{u}\|_2^2$; \quad $\mathcal{L}_p \mathrel{+}= \|\mathcal{F}_{\text{phy}}(\tilde{\bm{u}}, \bm{x}_t; \bm{\lambda})\|_2^2$
        \ENDFOR
    \ENDFOR
    \STATE $\mathcal{L} \gets \tfrac{1}{K}\sum_k\!\big(w_d \tfrac{\mathcal{L}_d}{|\mathcal{B}|} + w_p \tfrac{\mathcal{L}_p}{|\mathcal{B}|}\big) + \tfrac{1}{|\mathcal{D}|}\mathrm{KL}(Q(\bm{\theta};\bm{\zeta}) \| P(\bm{\theta}))$
    \STATE $\bm{\zeta}, \bm{\Theta}_{\text{DPC}} \gets \textsc{Adam}(\bm{\zeta}, \bm{\Theta}_{\text{DPC}}, \nabla\mathcal{L}, \eta)$
\ENDFOR
 
\Statex \textbf{Phase III: Inference}
\FOR{$s = 1$ to $S$}
    \STATE Sample $\bm{\theta}^{(s)} {\sim}\, Q(\bm{\theta}; \bm{\zeta}^*)$; \quad $\tilde{\bm{u}}^{(s)} \gets \textsc{ForwardPass}(\bm{x}_t, \bm{x}_{t-1}, \bm{\theta}^{(s)}, \bm{\Theta}_{\text{DPC}}^*)$
\ENDFOR
\RETURN $\hat{\bm{\mu}} = \tfrac{1}{S}\sum_s \tilde{\bm{u}}^{(s)}$, \; $\hat{\bm{\sigma}}^2 = \tfrac{1}{S}\sum_s (\tilde{\bm{u}}^{(s)} {-} \hat{\bm{\mu}})^2$
 
\end{algorithmic}
\end{algorithm}

\begin{figure*}[ht!]
    \centering
    \includegraphics[width=0.96\linewidth]{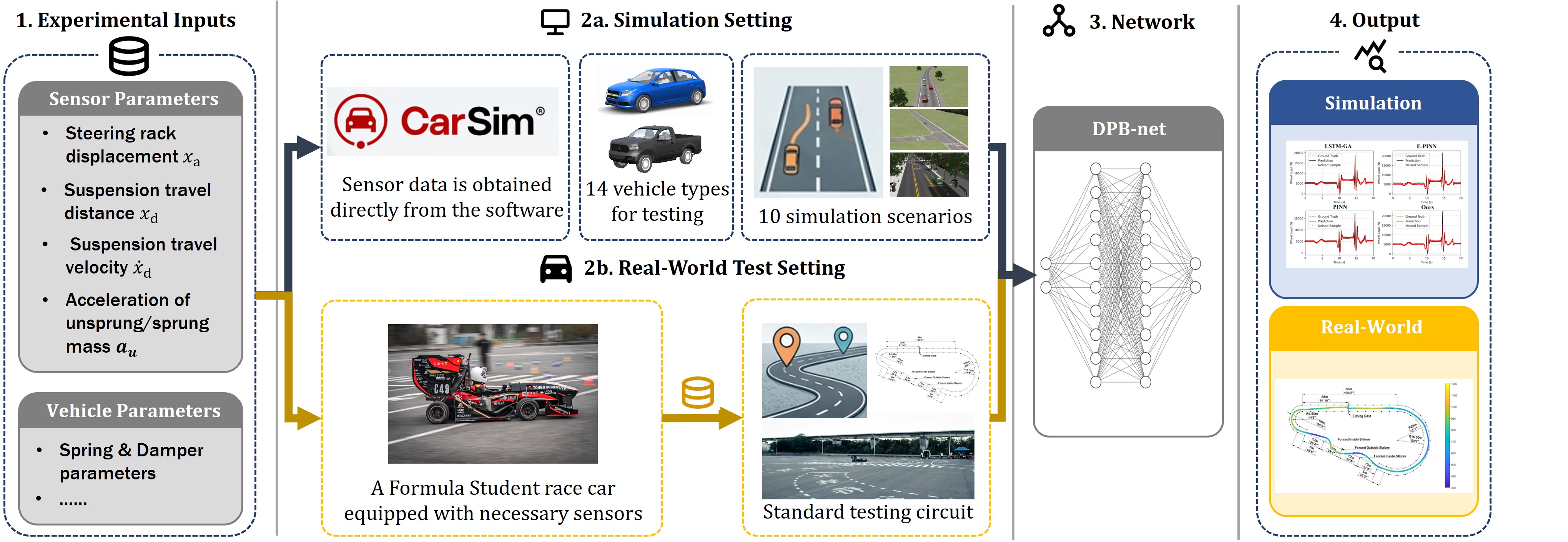}
    \caption{Overview of the experiments: In the simulation, the required input data are directly obtained by software and 10 types of vehicle and 4 types of testing scenarios are selected for data collection, while in the real-world experiments, input data are collected by sensors installed on the chassis and experiments are conducted on a Formula Student race car and completed in a standard test site. The wheel load is then calculated by the proposed \textbf{DBPnet} method.}
    \label{fig:expframe}
\end{figure*}

\section{Experiments}
\label{experiment}

Our experiments are conducted using the CarSim simulation software and a Formula Student race car, with the experimental framework shown in Figure \ref{fig:expframe}.

\subsection{Experimental Setup}

\subsubsection{Simulation Platform}

The simulation platform used is CarSim 2019.0 software in MATLAB/Simulink, a widely adopted vehicle dynamics simulation environment that features high-fidelity chassis structural interactions and supports a rich set of driving scenarios, ensuring that comparative experiments are conducted under strictly identical conditions.

\subsubsection{Hardware Platform}

These experiments are conducted on a Formula Student
racing car, and in order to make offline debugging more
convenient, the authors build a quarter suspension model,
as shown in Figure 7. 
The suspension model is equipped
with pushrod force sensor, knuckle acceleration sensor,
and suspension linear displacement sensor. However, due
to structural limitations, the steering wheel angle sensor
is not installed on the model, but the actual vehicle is
equipped with angle sensors and other necessary sensors
for vehicle operation.
Some related design parameters of the test vehicle
are shown in Table \ref{race}.

\begin{figure}
    \centering
    \includegraphics[width=0.95\linewidth]{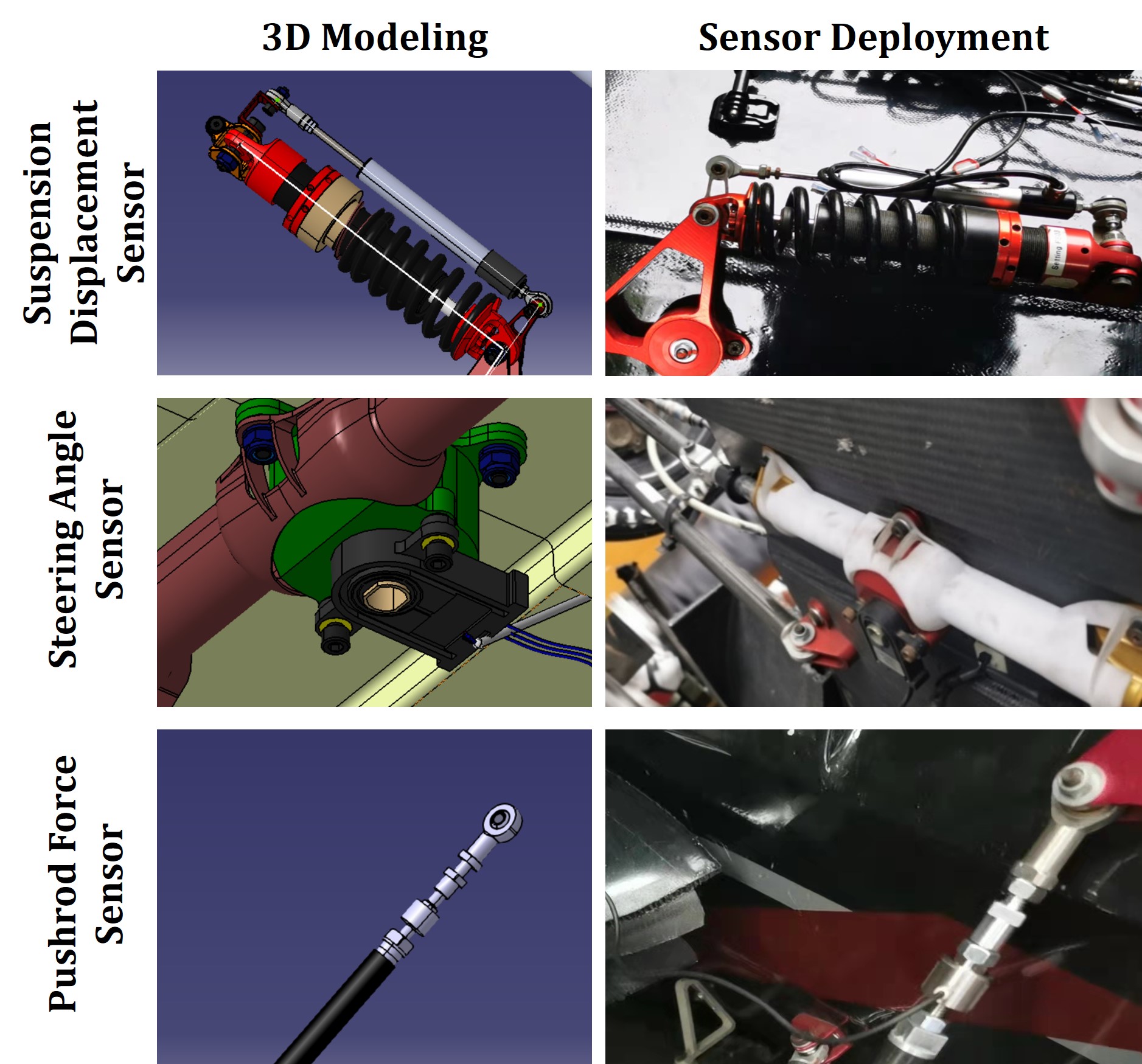}
    \caption{Design and deployment of several critical sensors are as follows:  
The suspension displacement sensor is mounted in parallel with the spring \& damper to monitor suspension compression, enabling the calculation of suspension geometry.  
The steering wheel angle sensor is installed in series at the end of the steering column.  
The suspension force sensor is integrated in series within the suspension linkage, connected via double-ended studs.}
    \label{fig:placeholder}
\end{figure}

\subsubsection{Baseline Methods}
We compare our method against multiple baselines, including the classical \textbf{EKF} \cite{ribeiro2004kalman} for state estimation, the widely used \textbf{LSTM} \cite{ip2021vehicle} for time-series prediction, and several PINN-based approaches including  \textbf{B-PINN} \cite{B-PINNs}, \textbf{A-PINN} \cite{Nabian2020} and \textbf{PINN} \cite{Guan2024}, thereby thoroughly validating the effectiveness of our method.

\subsection{Simulation Results}

\begin{figure*}[ht!]
\centering
\includegraphics[width=1\linewidth]{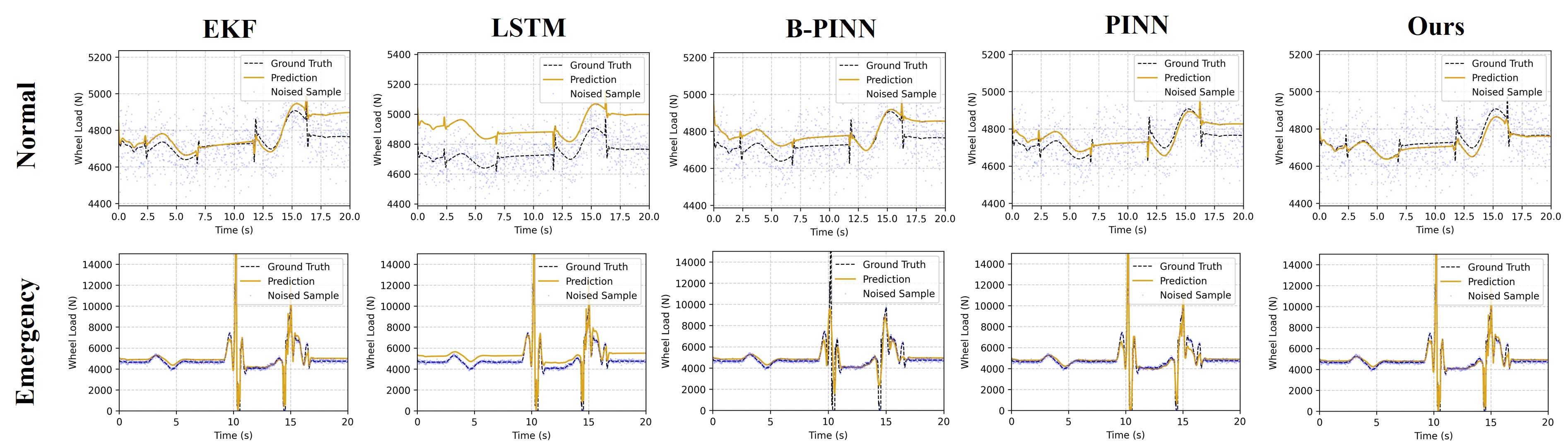}
\caption{Results of simulation: We select an urban driving scenario from the \textit{Normal Driving} condition and a distracted-driving scenario from the \textit{Emergency Driving} condition for illustration. As shown, our \textbf{DBPnet} method achieves the best performance.}
\label{fig:simres}
\end{figure*}

\begin{figure*}[ht!]
    \centering
    \includegraphics[width=\linewidth]{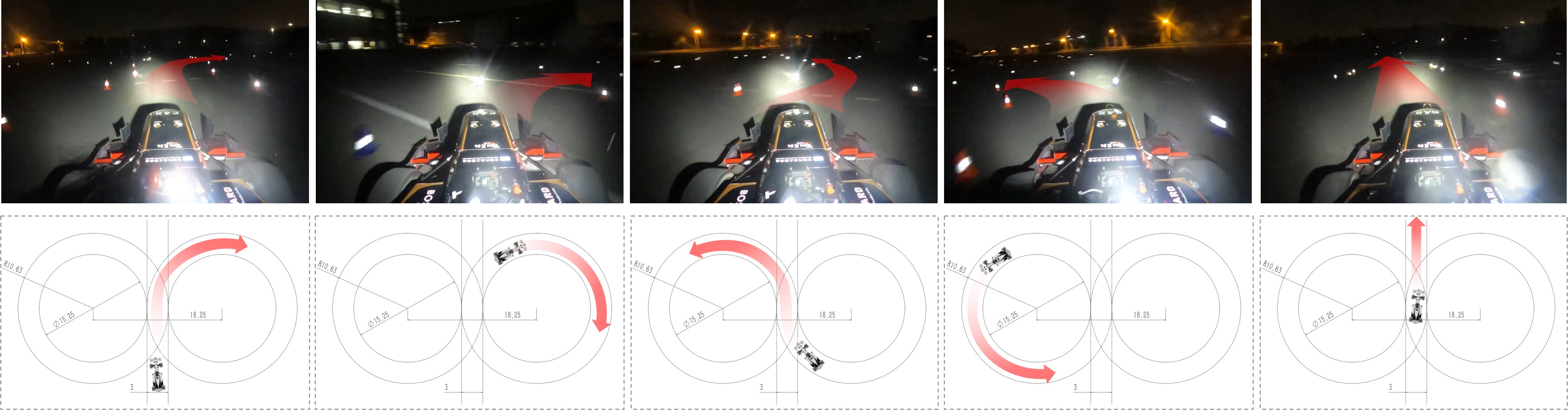}
    \caption{Real-world test of  ``figure-$8$" circuit: This is a standard race track with an inner diameter of $15.25$m, an outer diameter of $21.25$ m, a distance of $18.25$ m between the centers of the circles, and a track width of $3$ m. Race cars are required to drive two clockwise laps and then two counterclockwise laps within the track, and finally exit the track.}
    \label{fig:real}
\end{figure*}

\begin{figure*}[ht!]
    \centering
    \includegraphics[width=1\linewidth]{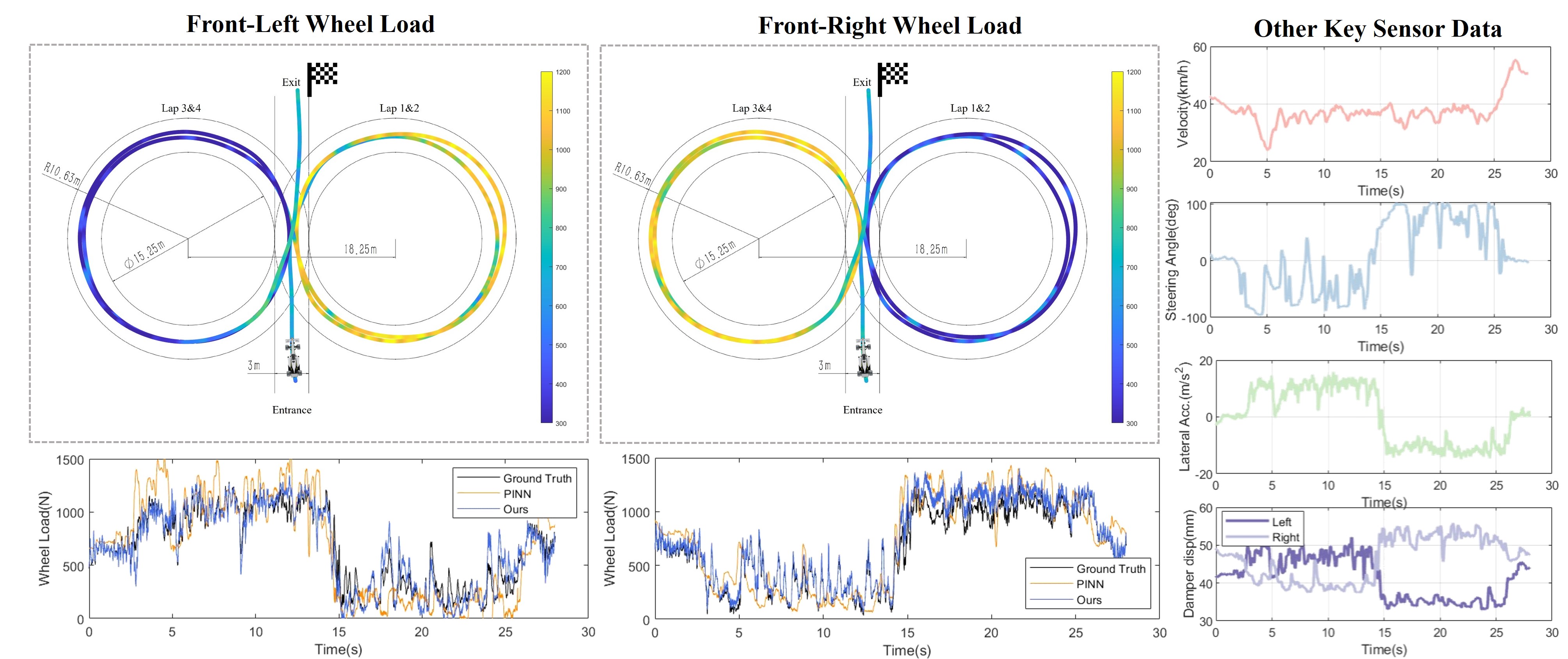}
    \caption{Results of real-world experiments: The top-left pair of graphs illustrate the distributions of front-left and front-right wheel loads; the bottom-left two graphs make a comparison between the wheel load values calculated by our \textbf{DBPnet} approach and those obtained via \textbf{PINN}; while the right-hand side presents information collected from the vehicle's speed sensor, steering angle sensor, acceleration sensor, and damper displacement sensor.}
    \label{fig:res2}
\end{figure*}


We divide the 10 selected simulation scenarios into two groups: \textit{Normal Driving} and \textit{Emergency Driving}, to evaluate method performance under distinct data distributions.  
\textit{Normal Driving} encompasses diverse settings such as urban roads, rural areas, and highways, where vehicle motion remains generally stable with minimal load variations.  
\textit{Emergency Driving} include hazardous scenarios like emergency avoidance and distracted driving, which involve abrupt braking, sharp steering, and other aggressive actions that induce severe load fluctuations and even risk vehicle rollover.

The quantitative results are presented in Table \ref{sample-table}.
We select one urban scenario from \textit{Normal Driving} and one distracted-driving emergency intervention scenario from \textit{Emergency Driving} for illustration in Figure \ref{fig:simres}.  

\begin{table}[t!]
\caption{Parameters of the Formula Student race car.}
\begin{center}
\setlength{\tabcolsep}{3pt}
\resizebox{1\linewidth}{!}{
\begin{tabular}{cccc}
\toprule
\textbf{Parameter} & \textbf{Value}& \textbf{Unit}& \textbf{Note} \\
\midrule
Sprung Mass& 248 & kg & With Driver\\

Unsprung Mass& 9 & kg & Single Wheel System\\
Wheel Base& 1600 & mm & -\\
Wheel Track Front& 1240 & mm & -\\
Wheel Track Rear& 1234 & mm & -\\
CG Height& 320 & mm & Center of Gravity\\
CG to Front Axle& 862 & mm & -\\
Suspension Stiffness Front& 78 & N/mm & -\\
Suspension Stiffness Rear& 100 & N/mm & -\\
Damper Front& 0.5-4.5 & kN$\cdot{}$s/m & Non-Linear\\
Damper Rear& 0.5-4.5 & kN$\cdot{}$s/m & Non-Linear\\

\bottomrule
\end{tabular}
}
\label{race}
\end{center}
\end{table}

\begin{table}[t!]
\caption{Simulation results of force prediction error (N).}
\label{sample-table}

\begin{center}
\resizebox{\linewidth}{!}{
\begin{tabular}{l|c|c}
\toprule
\textbf{Methods} & \textbf{Normal Driving} $\downarrow$ & \textbf{Emergency Driving} $\downarrow$\\
\midrule
EKF    & 259.751$\pm$20.981 &  1014.382$\pm$90.654\\
LSTM    &  308.512$\pm$33.217 &   1157.089$\pm$96.509\\
B-PINN    &  178.509$\pm$18.197 &   846.662$\pm$75.691\\
A-PINN    &  186.412$\pm$16.642 &   913.296$\pm$81.103\\
PINN   &  \underline{157.351$\pm$16.988} &   \underline{743.170$\pm$64.529}\\
\textbf{DBPnet (Ours)}    &  \textbf{105.886$\pm$10.674} &   \textbf{606.459$\pm$60.754}\\

\bottomrule
\multicolumn{3}{p{0.4\textwidth}}{\footnotesize Note: The \textbf{bold} data is the best one and the \underline{underlined} data has the second best performance.}
\end{tabular}
}
\end{center}
\end{table}

In \textit{Normal Driving}, all methods maintain errors within a relatively small range. 
The EKF, as a classical state estimation approach, demonstrates stable performance; however, due to its reliance on a fixed physical model and inability to capture data distribution characteristics from the dataset, it exhibits a comparatively larger bias. 
The LSTM, which primarily learns temporal patterns without encoding underlying physical principles, yields the largest error among all baselines. 
Several PINN-based methods perform well, benefiting from their explicit incorporation of physical constraints. 
Our proposed \textbf{DBPnet} method achieves the best performance by integrating Bayesian denoising with deep conditional embedding. 
As shown in the figure, our approach not only attains the smallest error during steady-state driving but also accurately captures fine-grained details of load oscillations.

In \textit{Emergency Driving}, the vehicle load undergoes drastic changes due to extreme maneuvers.
We present a distracted driving scenario, where the vehicle veers toward the roadside, begins to roll over, and is then urgently recovered by the driver.
Under such conditions, fixed physical models, typically derived under assumptions of stable vehicle operation, may no longer accurately represent the true dynamics. 
Consequently, the EKF incurs particularly large errors, and the LSTM also suffers from significant inaccuracies. 
Among the compared approaches, PINN-based methods yield relatively smaller errors. 
However, as shown in Figure \ref{fig:simres}, B-PINN exhibits poor fidelity in reconstructing sharp peaks, and the standard PINN also shows noticeable deviations in regions with high-amplitude oscillations. 
In summary, our \textbf{DBPnet} method achieves the best overall performance by deeply integrating the strengths of both physical priors and data-driven learning.

\begin{table}[t!]
\caption{Parameters of the sensors.}
\begin{center}
\setlength{\tabcolsep}{1pt}
\resizebox{1\linewidth}{!}{
\begin{tabular}{cccc}
\toprule
\textbf{Name} & \textbf{Frequency}& \textbf{Range}& \textbf{Note} \\
\midrule
Force Sensor& 20 Hz & $5\sim100$ (kg) & Suspension Force\\
Damper Disp. Sensor& 20 Hz & $0\sim110$ (mm) & Spring \& Damper\\
Steering Angle Sensor& 20 Hz & $-180 \sim180$ (deg) & -\\
Acceleration Sensor& 20 Hz & $10$ (g) & Any Direction\\
Velocity Sensor& 20 Hz & - & Wheel Velocity\\

\bottomrule
\end{tabular}
}
\label{tab:comp}
\end{center}
\end{table}

\subsection{Real-World Experiment}

The real-world experiments are conducted on a Formula Student race car, whose basic parameters are shown in Table \ref{race}. 
Additionally, we provide the data of the key sensors used in the experiments in Table \ref{tab:comp}.
It should be noted that the sensor frequencies listed in the table refer to the operating frequencies on our testing vehicle. Since velocity sensors typically do not specify a measurement range, and the vehicle speeds encountered during our real-world test are well within the operational limits of these sensors, the measurement range for the velocity sensor is not included in the table.
We use an Intel NUC as the in-vehicle computing unit, featuring an Intel 12th Gen Core i7-12700H CPU and an Intel Arc A770M 16GB GPU.
The ground truth is obtained by externally mounted transducers \cite{wang2019estimation}. 

\begin{figure*}[t!]
    \centering
    \includegraphics[width=\linewidth]{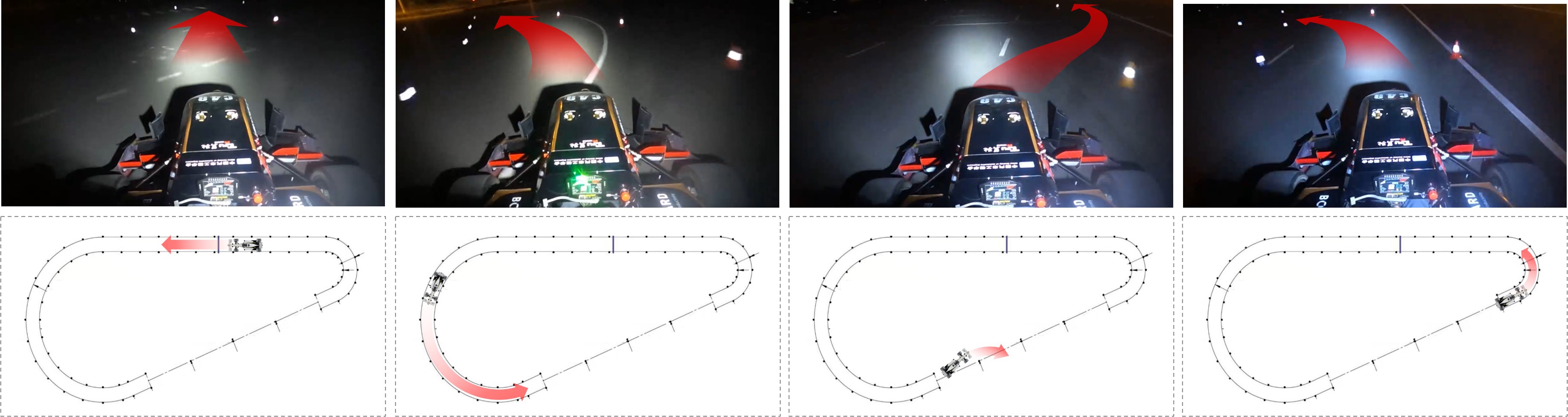}
    \caption{Real-world test of  ``asymmetric ellipse" circuit: This circuit includes high-speed corners, low-speed corners, long straightaway, and a slalom section, making it a minimal yet comprehensive standard scenario for evaluating the dynamic performance of race cars.}
    \label{fig:real2}
\end{figure*}

\begin{figure*}[t!]
    \centering
    \includegraphics[width=1\linewidth]{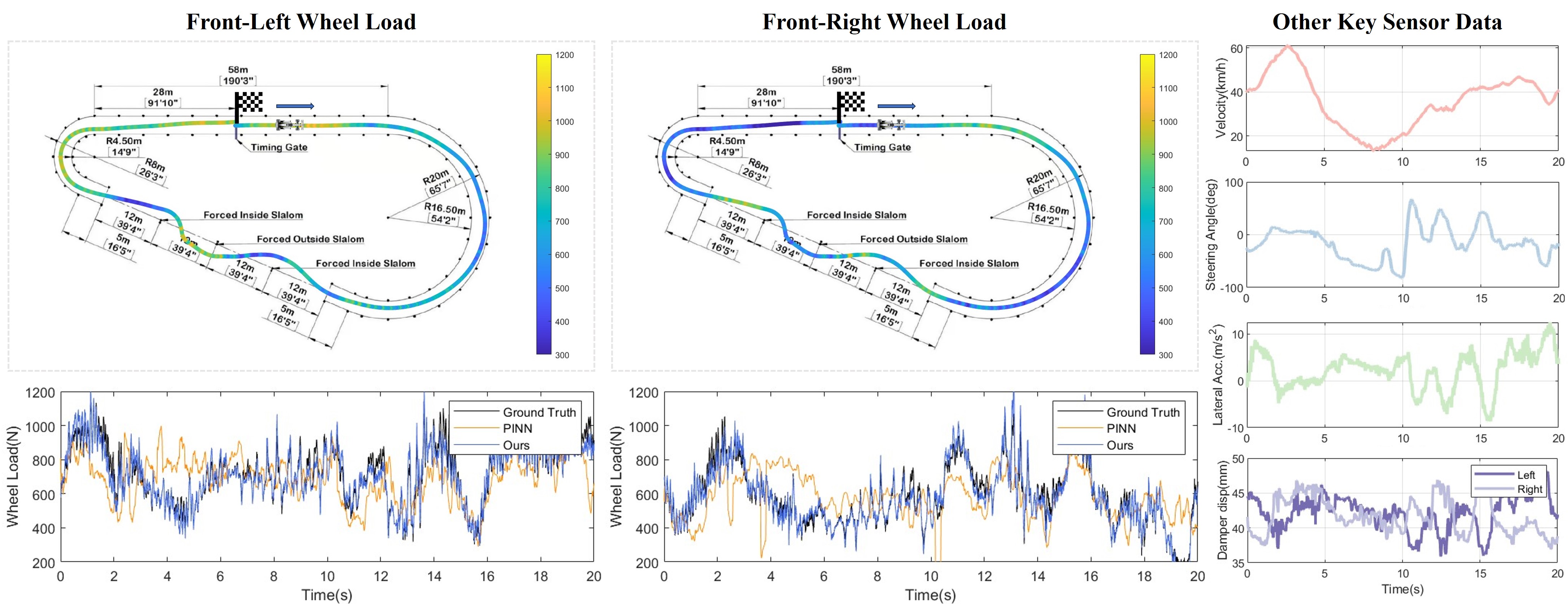}
    \caption{Results of real-world experiments: The top-left pair of graphs illustrate the distributions of front-left and front-right wheel loads; the bottom-left two graphs make a comparison. In our actual testing, for the convenience of track setup, we mirrored the layout of the real track, which is noted here for clarity.}
    \label{fig:res3}
\end{figure*}

The working conditions of the race car are extremely harsh: its chassis sensors are subject to significant noise interference, and the load also experiences large vibrations. The test condition that we select is a ``\textbf{figure-8}'' loop test, which consists of two clockwise laps and two counterclockwise laps, and a ``\textbf{asymmetric ellipse}'' test, which includes high-speed corners, low-speed corners, straightaways, and slalom sections, enabling a comprehensive test of load changes under steady-state and transient conditions.
The circuit are shown in Figure \ref{fig:real} and Figure \ref{fig:real2}.
Multiple sets of experiments were conducted, accumulating nearly $10$ hours in total, and two representative sets are selected here for demonstration, which are shown in Figure \ref{fig:res2} and \ref{fig:res3}.

As can be observed, the changing trend of wheel load distribution along the track is highly pronounced when steering. Moreover, combined with sensor data, wheel load fluctuations occur when the driver corrects the steering, which is reflected in the variations in the intensity of wheel load distributions. 
Additionally, since the static load of the vehicle is not symmetric between the left and right sides, the static positions of the left and right damper displacement sensors differ slightly. 
This also demonstrates the advantage of our \textbf{DBPnet} method, which enables wheel load calculation based on suspension characteristics rather than the overall vehicle characteristics.

\begin{table}[t!]
\caption{Real-world experiment results of force prediction error (N).}
\label{real}

\begin{center}
\begin{small}
\resizebox{\linewidth}{!}{
\begin{tabular}{l|c|c}
\toprule
\textbf{Method} & \textbf{Figure-8 $\downarrow$} & \textbf{Asymmetric Ellipse $\downarrow$}\\
\midrule

PINN    &  329.792$\pm$50.630 & 479.563$\pm$71.298 \\
\textbf{DBPnet} (Ours)    &  \textbf{194.586$\pm$39.094} & \textbf{255.049$\pm$46.961} \\

\bottomrule
\multicolumn{3}{p{0.4\textwidth}}{\footnotesize Note: The \textbf{bold} data is the best performance.}
\end{tabular}
}
\end{small}
\end{center}

\end{table}

The quantitative experimental results are presented in Table \ref{real}.
It can be observed that our \textbf{DBPnet} method, through the deep embedding of physical information and Bayesian operations, yields smaller errors compared to the other method.
It can also be observed that, due to the ``\textbf{asymmetric ellipse}'' condition involving more frequent steering changes and thus representing a non-steady-state condition, the overall error is slightly larger.
Moreover, it should be noted that the error data from real-vehicle testing are smaller than those from passenger car simulations, as the race car used in the actual tests has a very low mass, a difference inherently determined by the vehicle characteristics.

\begin{table}[t!]
\caption{Ablation study results of force prediction error (N).}
\label{ablation1}
\begin{center}
\resizebox{\linewidth}{!}{
\begin{tabular}{l|c|c}
\toprule
\textbf{Method} & \textbf{Normal Driving $\downarrow$} & \textbf{Emergency Driving $\downarrow$}\\
\midrule

w/o Physics Loss & 147.916$\pm$14.523 & 841.873$\pm$78.362 \\
w/o Bayesian     & 185.570$\pm$19.841 & 905.051$\pm$85.217 \\
w/o DPC          & 152.837$\pm$15.106 & 819.630$\pm$74.583 \\
\textbf{Full Model}  &  \textbf{105.886$\pm$10.674} &   \textbf{606.459$\pm$60.754}\\

\bottomrule
\multicolumn{3}{p{0.4\textwidth}}{\footnotesize Note: The \textbf{bold} data is the best performance.}
\end{tabular}
}
\vspace{-1pt}
\end{center}
\end{table}

\begin{table}[t!]
\caption{Running time and memory usage.}
\label{ablation2}
\begin{center}
\resizebox{\linewidth}{!}{
\begin{tabular}{l|c|c|c}
\toprule
\textbf{Setting} & \textbf{Train. Time} & \textbf{Inf. Time} & \textbf{Train. Memory}\\
\midrule

Simulation    &  50 min & 0.07 s & 2.5 GB\\
In-Vehicle    &  - & 0.10 s & - \\

\bottomrule
\end{tabular}
}
\vspace{-1pt}
\end{center}
\end{table}

\subsection{Ablation Study}
We conduct ablation experiments on several core modules, and the quantitative results are presented in Table \ref{ablation1}.
Ablation experiments verify the necessity of physical information loss, which ensures that the network output conforms to physical laws. 
In addition, the deep embedding of physical observation information by DPC also guides the network to fully learn the physical model and make real-time adjustments according to the observation information. 
Bayesian operation ensures that the network has robust output under noisy inputs. 
Ablation study demonstrates the effectiveness of each module proposed.


\subsection{Running Time and Memory}
We provide the model's training time, GPU memory usage, and inference time in Table \ref{ablation2}. 
The inference time refers to the time per sample, not the total time.
Since our method does not show a clear advantage over other methods in terms of runtime and memory, we do not report the timing and memory usage metrics for those other methods. 
However, we present the data from our own method obtained in both simulation and real-vehicle tests.
Since the sensor sampling frequency is higher than the onboard computation frequency, the frequency used in our vehicle data is the computation frequency.

\section{Conclusion and Future Work}
\label{conclusion}

In this paper, we propose a Damper characteristics-based Bayesian Physics-informed neural network (\textbf{DBPnet}) for state estimation in MIMO systems and validate its effectiveness in a vehicle with ADAS.
To address the challenge of training in MIMO systems due to the noise and imperfect correspondence between inputs and outputs, we introduce the Bayesian operation into PINN, where the posterior distribution of parameters is derived by using the VI method, ensuring improved uncertainty quantification. 
Furthermore, a physics-informed loss function is integrated with the data loss to ensure that the network outputs adhere to physical laws while preventing overfitting.
Moreover, a physical conditioning embedding module inspired by damper characteristic is proposed, which extracts the temporal features of the input signals and embeds the physics guidance into each layer of the PINN, preventing the model from misguided by a fixed physical model.
Results of simulations and real-world experiments demonstrate that the proposed \textbf{DBPnet} consistently achieves high accuracy and has robust perfomance in most cases. 
However, the proposed \textbf{DBPnet} also has certain limitations. Due to the characteristics of the network structure, its inference speed is slower than that of other baseline networks. In future work, we will focus on addressing the issue of inference speed and attempt to validate the method in more engineering systems.

\section*{ACKNOWLEDGMENTS}
This research was supported by the National Nature Science Foundation of China (No. 52432014 and No. 52275123).
Tianyi Wang, Xiangyu Li, Yiming Xu, Sikai Chen, Junfeng Jiao, and Christian Claudel did not receive any funding for this work.

\bibliographystyle{IEEEtran}
\bibliography{references}

\newpage

\section{Biography}
 



\begin{IEEEbiographynophoto}{Tianyi Wang} is currently a Ph.D. student at the Department of Civil, Architectural, and Environmental Engineering, The University of Texas at Austin, TX, USA. 
He received his M.S. degree in mechanical engineering and materials science from Yale University, CT, USA, in 2025. 
He earned his B.E. degree in vehicle engineering from Tongji University, Shanghai, China, in 2024. 
His research interests include intelligent transportation systems and connected and automated vehicles, particularly in decision-making, trajectory-planning and cooperative control. 
\end{IEEEbiographynophoto}

\begin{IEEEbiographynophoto}{Tianyi Zeng} is currently a M.E. student at the School of Automation and Intelligent Sensing, Shanghai Jiao Tong University, Shanghai, China.
He received his B.E. degree in vehicle engineering from Tongji University, Shanghai, China, in 2024. 
His research interests include generation models, physics-informed learning, large language models and intelligent vehicles. 
\end{IEEEbiographynophoto}

\begin{IEEEbiographynophoto}{Zimo Zeng} is currently a M.E. student at the College of Electrical Engineering, Zhejiang University, Zhejiang, China.
She received her B.E. degree in vehicle engineering from Tongji University, Shanghai, China, in 2025. 
Her research interests include world models, vision-language-action models and diffusion models. 
\end{IEEEbiographynophoto}

\begin{IEEEbiographynophoto}{Feiyang Zhang} is currently a M.E. student at the School of Automotive Studies, Tongji University, Shanghai, China.
He received his B.E. degree in vehicle engineering from Tongji University, Shanghai, China, in 2025. 
His research interests include autonomous vehicles, advanced control and world models. 
\end{IEEEbiographynophoto}

\begin{IEEEbiographynophoto}{Yujin Wang} is currently a Ph.D. student at the School of Automotive Studies, Tongji University, Shanghai, China.
He received his B.E. degree in vehicle engineering from Tongji University, Shanghai, China, in 2023.
His research interests include vision-language models, deep learning, and foundation intelligence in autonomous driving. 
\end{IEEEbiographynophoto}

\begin{IEEEbiographynophoto}{Xiangyu Li} is currently a Ph.D. student at the Department of Civil, Architectural, and Environmental Engineering, The University of Texas at Austin. 
He received the M.S. degree in computer engineering from Northwestern University in 2025, and the M.S. degree in transportation engineering from the University of California, Berkeley, in 2023. 
He earned the B.E. degree in transportation engineering from Beijing Jiaotong University in 2022. 
His research interests include autonomous driving, intelligent transportation systems, generative AI, and computer vision.
\end{IEEEbiographynophoto}

\begin{IEEEbiographynophoto}{Yiming Xu} received the Ph.D. degree in civil engineering from University of Florida in 2023. 
He is currently a Postdoctoral Fellow in the School of Architecture at the University of Texas at Austin. 
His work focuses on developing and applying machine learning methods to tackle challenges in transportation systems. 
He specializes in trustworthy machine learning and deep learning applications in travel behavior analysis and time series modeling to support urban mobility management and operation.
\end{IEEEbiographynophoto}

\begin{IEEEbiographynophoto}{Sikai Chen} received the Ph.D. degree in civil engineering, with a focus on computational science and engineering, from Purdue University in 2019. 
He is currently an Assistant Professor in the Department of Civil and Environmental Engineering and the Department of Mechanical Engineering (courtesy) at the University of Wisconsin–Madison.
He is a member of the American Society of Civil Engineers Connected and Autonomous Vehicle Impacts and Economics and Finance national committees, IEEE Emerging Transportation Technology Testing Technical Committee, and Transportation Research Board Standing Committee on Statistical Methods (AED60). 
He is a Senior Member of IEEE.
His research interests include human users, artificial intelligence, and transportation. 

\end{IEEEbiographynophoto}

\begin{IEEEbiographynophoto}{Junfeng Jiao} received the Ph.D. degree in urban planning from University of Washington in 2010.
He is currently an Associate Professor in the Community and Regional Planning Program at The University of Texas at Austin. 
He is the founding director of Urban Information Lab, director of Texas Smart Cities, director of UT Ethical AI program, and a founding member of UT Austin's Good Systems Grand Challenge. 
His research focuses on Smart Cities, Urban Informatics, and Ethical/Generative AI. 
\end{IEEEbiographynophoto}

\begin{IEEEbiographynophoto}{Christian Claudel} received the Ph.D. degree in electrical engineering from University of California Berkeley in 2010.
He is currently an Associate Professor in the Department of Civil, Architectural, and Environmental Engineering at The University of Texas at Austin. 
His research interests include control and estimation of distributed parameter systems, cyber-physical systems monitoring, and the use of wireless sensor networks for environmental applications.
\end{IEEEbiographynophoto}

\begin{IEEEbiographynophoto}{Xinbo Chen} received the Ph.D. degree in precision engineering from Tohoku University in 1995.
Since 2002, he has been a Professor with the School of Automotive Studies at Tongji University. 
From 2000 to 2001, he was a Visiting Researcher with the Tokyo Institute of Technology, Tokyo, Japan. 
From 2013 to 2014, he was a Visiting Professor with the Tokyo University of Agriculture and Technology, Tokyo, and Southern Illinois University at Edwardsville, Edwardsville, IL, USA. 
His current research interests include electric propulsion system of electric vehicle (EV), vehicle chassis electrification and coordinated dynamics control, transmission system optimal design, and active/semiactive suspension control.
\end{IEEEbiographynophoto}

\vfill

\end{document}